\documentclass[amsmath, amssymb,aps, pre,twocolumn]{revtex4-1}

\usepackage[utf8]{inputenc}
\usepackage{graphicx}
\usepackage{color}
\usepackage{bigints}

\newcommand{\eps}{\varepsilon}
\newcommand{\s}{\boldsymbol{s}}

\DeclareGraphicsExtensions{.pdf}
\graphicspath{{./figs/}}

\begin{document}
\pagestyle{plain}
\title{Topological bifurcations in a model society of reasonable contrarians}

\author{Franco Bagnoli}
\email{franco.bagnoli@unifi.it}
\affiliation{Dipartimento di Fisica e Astronomia,
 Universit\`a di Firenze,\\
 Via G. Sansone 1,  50017 Sesto Fiorentino (FI), Italy; \\also 
 INFN, sez.\ Firenze.}
\author{Ra\'ul Rechtman}
\email{rrs@cie.unam.mx}
\affiliation{Instituto de Energ\'i{}as Renovables, Universidad Nacional 
 Aut\'onoma de M\'exico,\\  Apdo.\ Postal 34, 62580 Temixco Mor., M\'exico}

\begin{abstract}
  People are often divided into conformists and contrarians, the
  former tending to align to the majority opinion in their
  neighborhood and the latter tending to disagree with that
  majority. In practice, however, the contrarian tendency is rarely
  followed when there is an overwhelming majority with a given
  opinion, which denotes a social norm. Such reasonable contrarian
  behavior is often considered a mark of independent thought, 
  and can be a useful strategy
  in financial markets.
  
  We present the opinion dynamics of a society
  of reasonable contrarian agents. The model is a
  cellular automaton of Ising type, with antiferromagnetic pair
  interactions modeling contrarianism and plaquette terms modeling
  social norms.  We introduce the entropy of the collective variable as 
  a way of comparing deterministic (mean-field) and probabilistic 
  (simulations) bifurcation diagrams.
  
  In the mean field approximation the model exhibits
  bifurcations and a chaotic phase, interpreted as coherent
  oscillations of the whole society. However, in a one-dimensional
  spatial arrangement one observes incoherent oscillations and a
  constant average. 
  
  In simulations on Watts-Strogatz networks with a
  small-world effect the mean field behavior is recovered, 
  with a bifurcation diagram that resembles the mean-field 
  one, but using the rewiring probability as the control parameter. 
   Similar bifurcation diagrams are found for scale free networks, 
   and we are able to compute an effective connectivity for such networks. 
   
\end{abstract}

\maketitle

\section{Introduction}

%

Social norms are the basis of a community. They are often adopted and
respected even if in contrast with an agent's immediate
advantage, or, alternatively, even if they are costly with respect to
a naive behavior. Indeed, the social pressure towards a widespread
social norm is sometimes more powerful than a norm imposed by
punishments.

It is well known that the establishment of social norms is a difficult task and their imposition is not always fulfilled. This problem has been affronted
by Axelrod in a game-theoretic formulation~\cite{Axelrod:EvolutionOfCooperation}, as the foundation of the cooperation and of the society itself. Axelrod's idea is that of a repeated game. Although in a one-shot game it is always profitable to win not following any norm, in a repeated game there might be several reasons for cooperation~\cite{Nowak:FiveRules}, the most common ones are direct reciprocity and reputation. In all these games, the crucial parameters are the cost of cooperation with respect to defeat, and the expected number of re-encounters with one's opponent or the probability that one's behavior will become public. One can assume that these aspects are related to the size of the local community with which one interacts and the fraction of people in this community that share the acceptance of the social norm. Indeed, the behavior of a spatial social game is strongly influenced by the network structure~\cite{Klemm}.

In the presence of a social norm, people can manifest a conformist or
a non conformist or contrarian attitude, characterized by the
propensity to agree or disagree with the average opinion in their
neighborhood.

Contrarian agents were first discussed in the field of finance~\cite{corcos02} and later in opinion formation models~\cite{galam04}. Contrarian behavior may have an advantage in financial investment.  Financial contrarians look for mispriced investments, buying those that appear to be undervalued by the market and selling those that are overpriced.  In opinion formation models, contrarians gather the average opinion of their neighbors and choose the opposite one. Reasonable contrarians do not violate social norms, \textit{i.e.}, they agree with conformists if the majority of neighbors is above a certain threshold. 

Models of social dynamics have been studied extensively.~\cite{review}.
In this paper we model the dynamics of a homogeneous community with
different degrees of reasonable contrarianism. 

One of the main motivations for this study is that of exploring the possible behavior of autonomous agents employed  in algorithmic trading in an electronic  market. Virtually all markets are now electronic~\cite{J} and the speed of transaction require the use of automatic agents (algorithmic trading)~\cite{AT}. 
Our study can be consider as an exploration of possible collective effects in a homogeneous automatic market.

We consider a simplified cellular automaton 
model~\cite{Bagnoli:ChaosUniformSociety}. 
Each agent can
have one of two opinions at time $t$, and we study the parallel
evolution of such agents, which can be seen also as a spin system.

A society of conformists can be modeled as a ferromagnet and one of
contrarians as an antiferromagnet.  Each agent changes his
opinion according to the local social pressure or what is the same,
the average opinion of his neighbors. 
In spin language, social norms can be represented as plaquette terms
since they are non additive and important when the social
pressure is above or below given thresholds.

The model, presented in Sec.~\ref{sec:model}, simulates a society of
$N$ reasonable contrarians that can express one of two opinions, 0 and 1. At
each time step, each agent changes its opinion according to a transition
probability that takes into account the average opinion of his
neighbors, that is, the local social pressure and the adherence to social
norms. The neighborhoods are fixed in time. The transition probability
depends on a parameter $J$, analogous to the spin coupling in the
Ising model, which is positive for a society of conformists
(ferromagnet) and negative for one of contrarians (antiferromagnet).

For a one dimensional society where the neighborhood of each agent
includes its first $k$ nearest neighbors, $k$ is the connectivity,
the average opinion 
fluctuates around the value 1/2, regardless of the values of the parameters of
the transition probability.  Simulations of the one-dimensional
version of the model show irregular fluctuations at the microscopic
level, with short range correlations~\cite{Bagnoli:longrange}.

The mean field approximation of the model for the average opinion 
is a discrete map which exhibits bifurcation diagrams as the
parameters $k$ and $J$ change, as discussed in
Sec.~\ref{sec:meanfield}. The diagrams show a period doubling route
towards chaos. 

In Sec.~\ref{sec:smallworld} we discuss the model on Watts-Strogatz networks that exhibit  
the small-world effect~\cite{WattsStrogatz}. We find a bifurcation
diagram as the fraction $p$ of rewired links changes. Since the opinion
of agents change probabilistically, we speak of probabilistic
bifurcation diagrams. 

In Sec.~\ref{sec:networks}, the reasonable nonconformist opinion model
is extended to scale-free networks. Again, we observe a probabilistic
bifurcation diagram, similar to the previous ones, by varying the
coupling $J$. We are able to obtain a good
mapping of the scale-free parameters onto the mean field approximation
with fixed connectivity $k$.

In order to compare the deterministic and probabilistic bifurcation diagrams, we exploit the entropy $\eta$ of the average opinion. In the deterministic case, large values of $\eta$ correspond to positive values of
the Lyapunov exponent. In Secs.~\ref{sec:meanfield}, \ref{sec:smallworld}, 
and \ref{sec:networks} we show that $\eta$ can be used to characterize
numerically order and disorder in deterministic and probabilistic bifurcation
diagrams. Finally we present some conclusions. 


\section{The model}
\label{sec:model}

Each of the $N$ agents  has
opinion $s_i(t)$ at the discrete time $t$ with $s_i\in\{0,1\}$ and
$i=0,\dots,N-1$. The state of the society is $\s=(s_0,\dots,s_{N-1})$.
In the context of cellular automata and discrete magnetic systems, the
state at site $i$ is $s_i$ and the spin at site $i$ is
$\sigma_i=2s_i-1$ respectively. The average opinion $c$ is given by
\begin{equation}
 \label{eq:c}
 c=\dfrac{1}{N}\sum_i s_i.
\end{equation}
\begin{figure}
 \begin{center}
  \includegraphics[width=0.8\columnwidth]{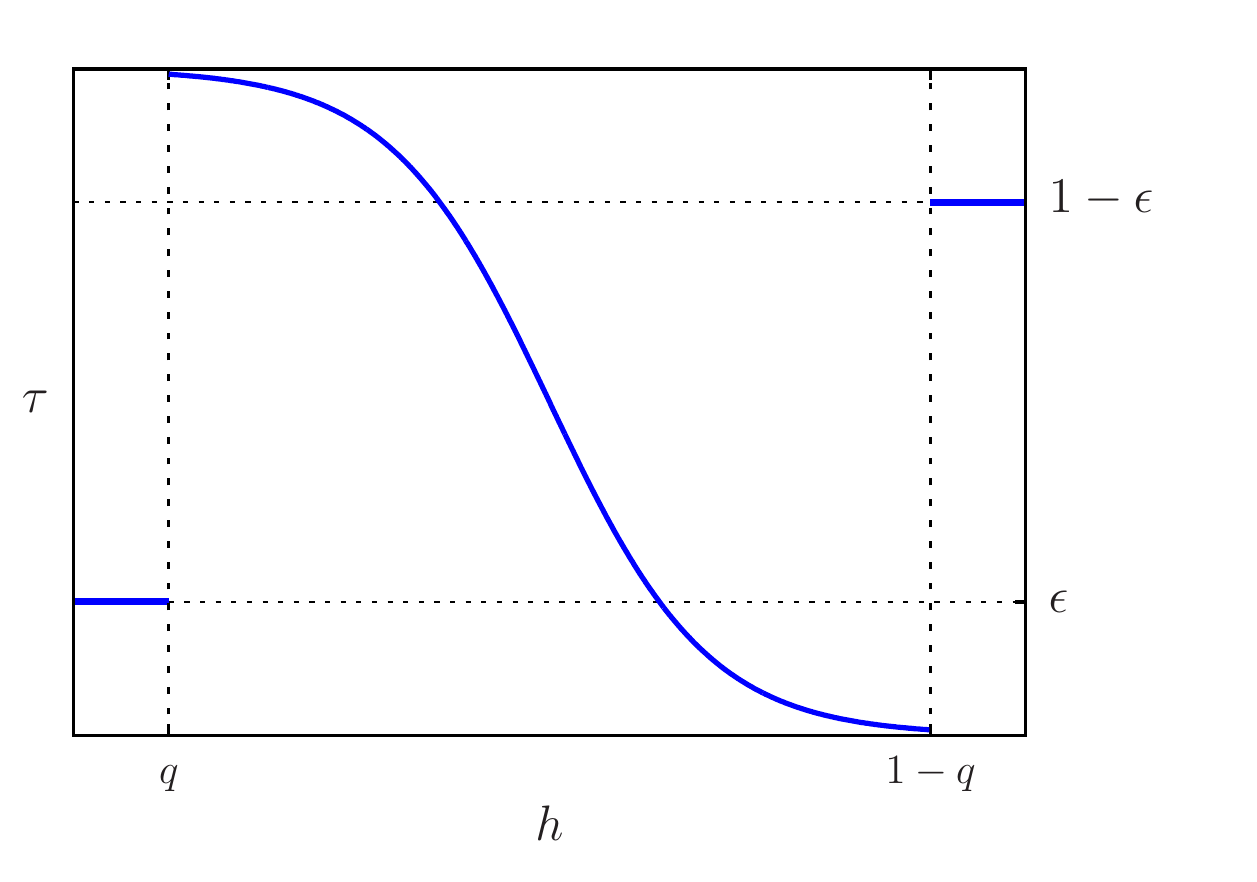}
 \end{center}
 \caption{\label{fig:tau-mf} (Color online)  The transition
    probability $\tau(h)$ given by Eq.~\eqref{tau} with $J=-3$,
    $k=20$, $q=0.1$, and $\eps = 0.2$.}
\end{figure}

The opinion of agent $i$ evolves in time according to the opinions of
his neighbors, identified by an adjacency matrix with components
$a_{ij}\in\{0,1\}$. If agent $j$ is a neighbor of agent $i$,
$a_{ij}=1$, otherwise $a_{ij}=0$. The adjacency matrix defines the
network of interactions and is considered fixed in time.  The
connectivity $k_i$ of agent $i$ is the size of his neighborhood,
\[
 k_i = \sum_j a_{ij}.
\]
The average local opinion or social pressure $h_i$, is defined by
\begin{equation}\label{h}
 h_i =\dfrac{\sum_{j} a_{ij}s_j}{k_i}.
\end{equation}

The opinion of agent $i$ changes in time according to the transition
probability $\tau(s_i|h_i)$ that agent $i$ will hold the opinion $s_i$
at time $t+1$ given the local opinion $h_i$ at time $t$. This
transition probability, shown in Fig.~\ref{fig:tau-mf}, is given by
\begin{equation}\label{tau}
 \tau(h) =
  \begin{cases}
    \eps & \text{if $h<q$,}\\
    \dfrac{1}{1+\exp(-2J(2h-1))} & \text{if $q\le h\le 1-q$,}\\
    1-\eps & \text{if $h>1-q$,}
  \end{cases}
\end{equation}
with $\tau(h)=\tau(1|h)$.
The quantity $q$ denotes the threshold
for the social norm, and $\eps$ the probability of being reasonable.
With $\eps=0$ or $q=0$, $\s=(0,\dots,0)$ and $\s=(1,\dots,1)$ are
absorbing states~\cite{Bagnoli:longrange}. In the following we set
$\eps=0.2$ and $q=0.1$ if not otherwise stated. The results are qualitatively independent of
$\eps$ and $q$ as long as they are small and positive.
The transition probability $\tau$ has the symmetry
\begin{equation}
 \tau(1-h)=1-\tau(h).
\end{equation}

With $J>0$ and $q<h_i<1-q$, agent $i$ will likely agree with his neighbors, 
a society of conformists. With $J<0$ and $q<h_i<1-q$, agent $i$ will 
likely disagree with his neighbors, a contrarian society. For 
$0\leq h\leq q$ or
$1-q\leq h\leq 1$ agent $i$ will likely agree (if $\eps$ is small) with the majority of his neighbors, regardless of the value of $J$.
  
We might also add an external field $H$, modeling
news and broadcasting media, but in this study we always keep $H=0$.
We are thus modeling a completely uniform society, \emph{i.e.}, we
assume that the agent variations in the response to stimuli are
quite small. Moreover, we do not include any memory effect, so that
the dynamics is completely Markovian. 

In the language of spin systems, $\tau(h_i)$ is the transition
probability of the heat bath dynamics of a parallel Ising model with
ferromagnetic, $J>0$, or antiferromagnetic, $J<0$,
interactions~\cite{Derrida}. The behavior of the transition
probability in the regions $h<q$ and $h > 1-q$ may be seen as due to a
non-linear plaquette term that modifies the ferro/antiferro
interaction. If we set $\eps=0$ and $J=-\infty$, the system becomes
deterministic (in magnetic terms, this is the limit of zero
temperature). 
  
In one dimension, with $k=3$, $1/3<q\le 1/2$ and $\eps=0$ this model
exhibits a nontrivial phase diagram, with two directed-percolation
transition lines that meet a first-order transition line in a critical
point, belonging to the parity conservation universality
class~\cite{Bagnoli:3inputs}. In this case, we have the stability of
the two absorbing states for $J>0$ (conformist society or ordered
phase), while for $J<0$ (anti-ferro or contrarian) the absorbing
states are unstable and a new, disordered active phase is observed.
The model has been studied in the one-dimensional case with
larger neighborhood~\cite{Bagnoli:longrange}. In this case one observes
again the transition from an ordered to an active, microscopically
disordered  phase, but with no coherent oscillations. Indeed, if the system
enters  a truly disordered configuration, then the local field $h$ is
everywhere equal to $0.5$ and the transition probabilities $\tau$
become insensitive to $J$ and equal to $0.5$, see Eq.~\eqref{tau}. 


\section{Mean field approximation}
\label{sec:meanfield}

\begin{figure}
 \begin{center}
  \begin{tabular}{cc}
    (a) & (b) \\
   \includegraphics[width=0.45\columnwidth]{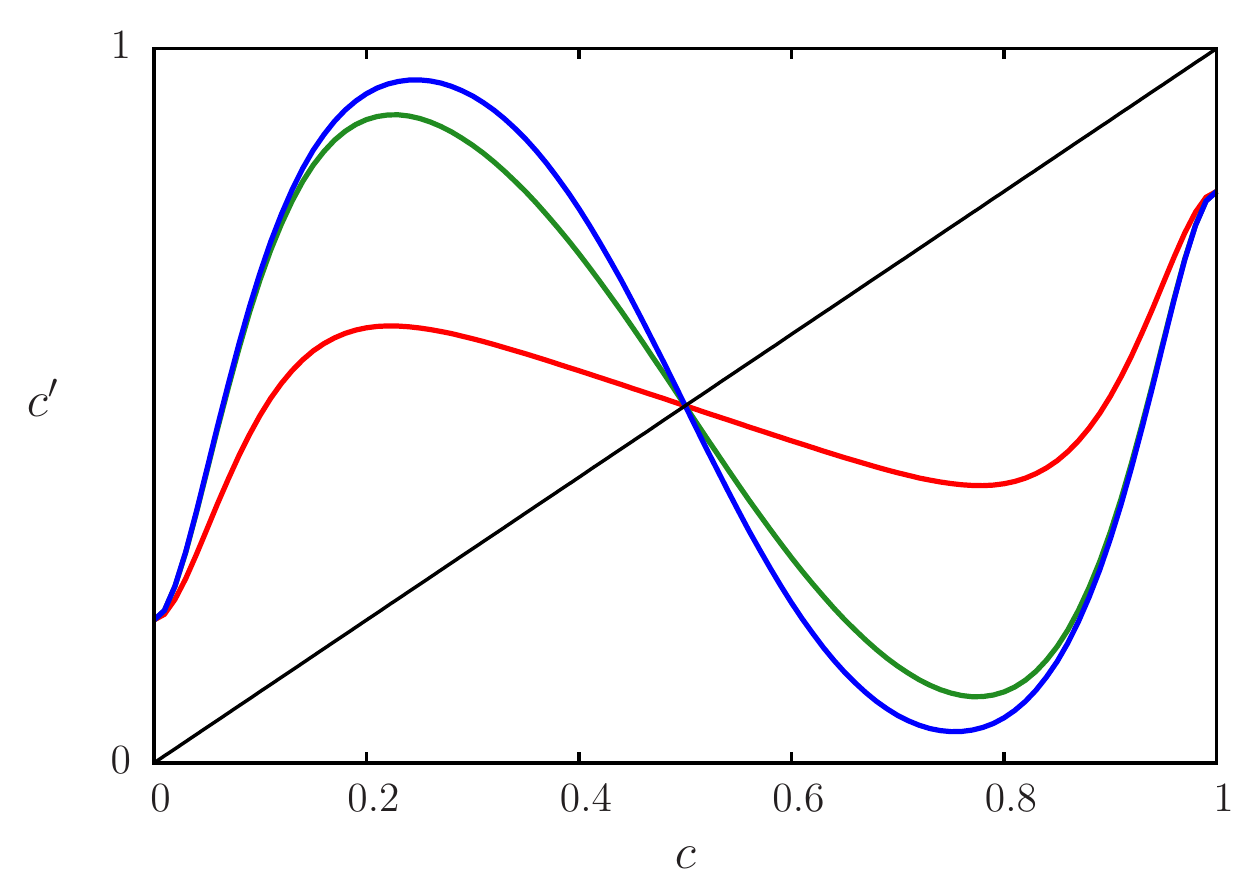} &
   \includegraphics[width=0.45\columnwidth]{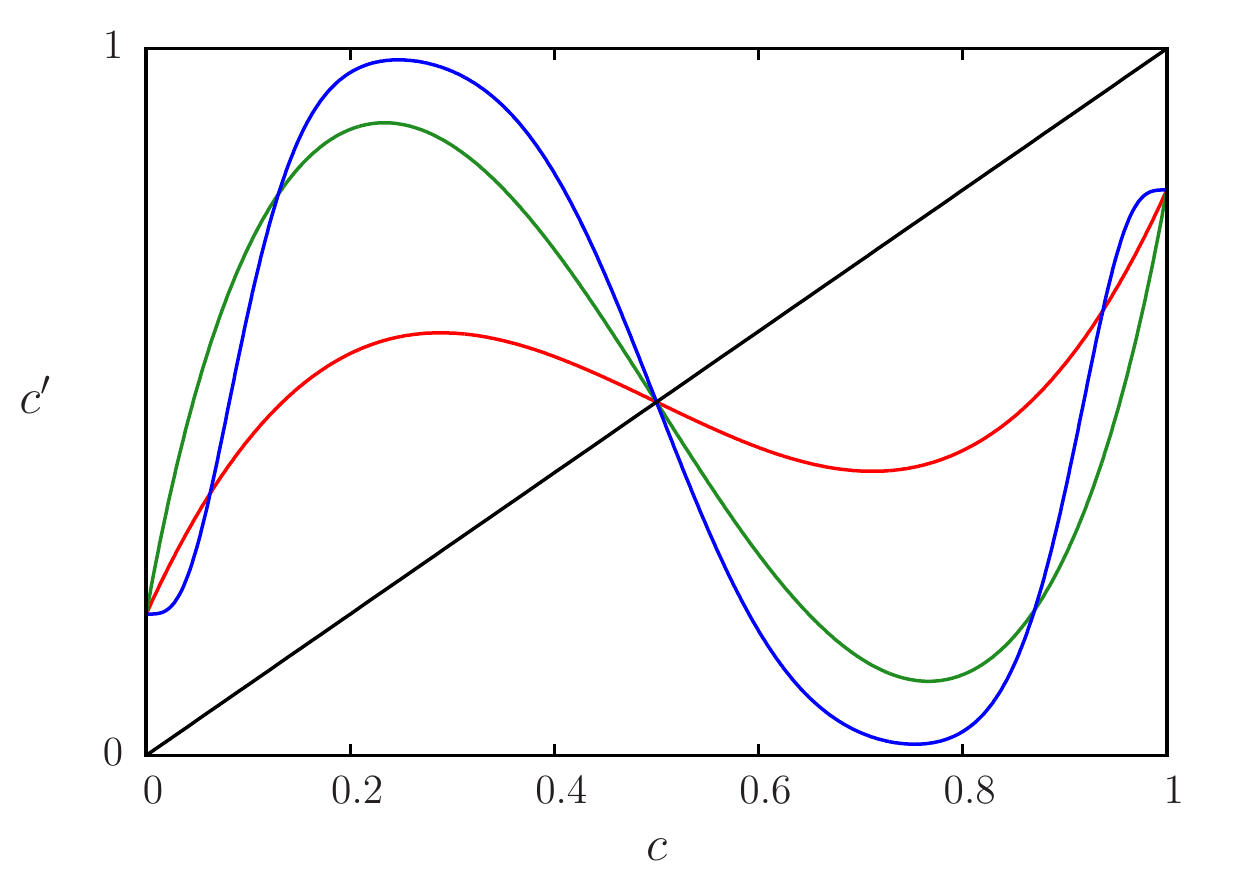} 
  \end{tabular}
 \end{center}
  \caption{\label{fig:mfeq} (Color online)  (a) Graphs of the mean
    field map, Eq.~(\ref{eq:mf}) for different values of $J$ and
    $k=20$. From bottom to top for
    $c<1/2$, $J=-0.5$ (red), $J=-3.0$ (green), and $J=-6.0$
    (blue). (b) Graphs of Eq.~(\ref{eq:mf}) for different values of $k$ and
    $J=-6$. From bottom to top for $c\sim 0.2$,
    $k=4$ (red), $k=10$ (green), and $k=38$ (blue).} 
\end{figure}

The simplest mean-field description of the model is given by 
\begin{equation}\label{eq:mf}
  c' = f(c)=\sum_{w=0}^k \binom{k}{w} c^w (1-c)^{k-w} \tau\left(w/k\right),
\end{equation}
with $c'=c(t+1)$ and $c=c(t)$~\cite{ilachinski}.  The term in 
parenthesis on the {\em r.h.s} of this expression denotes the 
$w$-combinations from a set of $k$ elements. In Fig.~\ref{fig:mfeq}
we show some graphs of $f$. The map $f$ has the same symmetry property as the
transition probabilities $\tau$,
\begin{equation}
 \label{eq:sym}
 f(1-c)=1-f(c).
\end{equation}

\begin{figure}
 \begin{center}
  (a)\\
  \hskip -11mm
   \includegraphics[width=0.84\columnwidth]{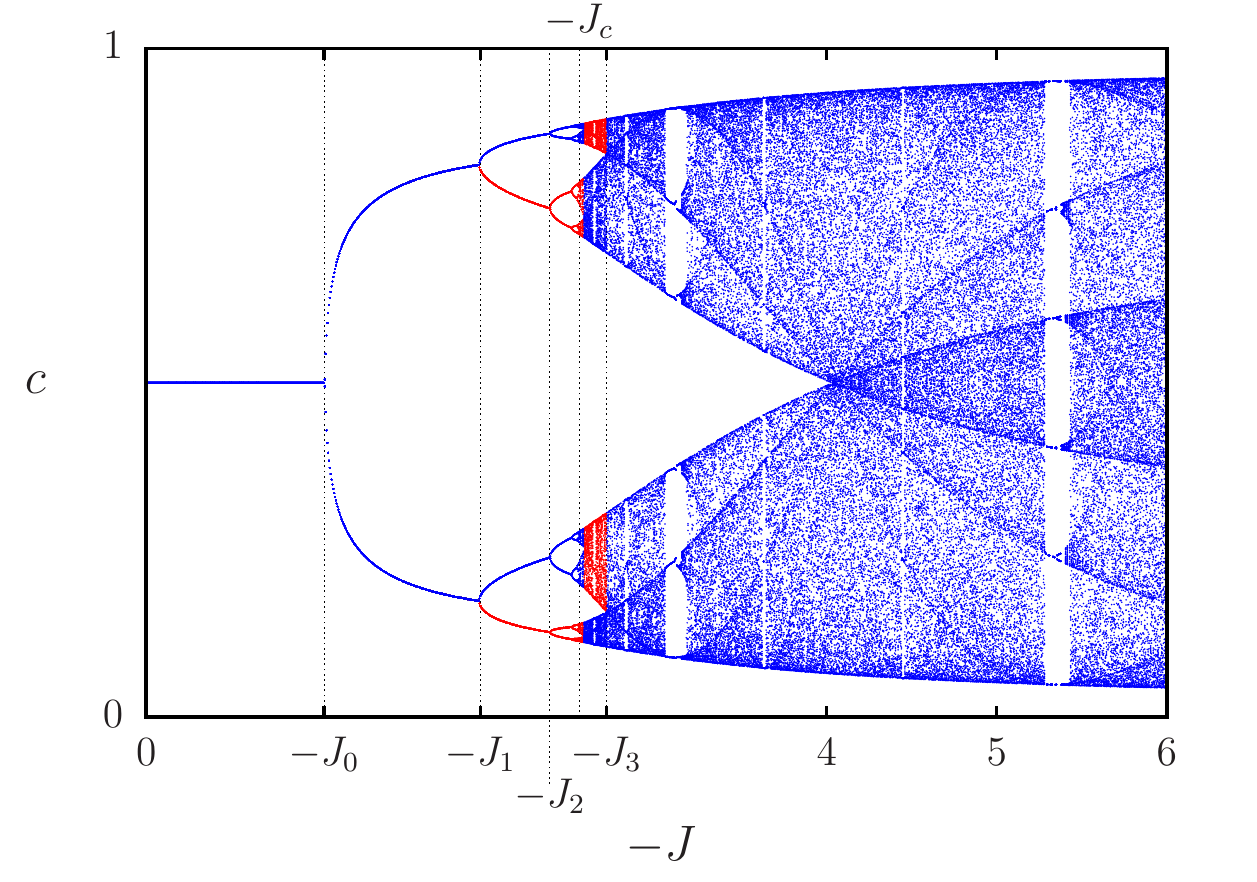}\\
  (b)
  \includegraphics[width=\columnwidth]{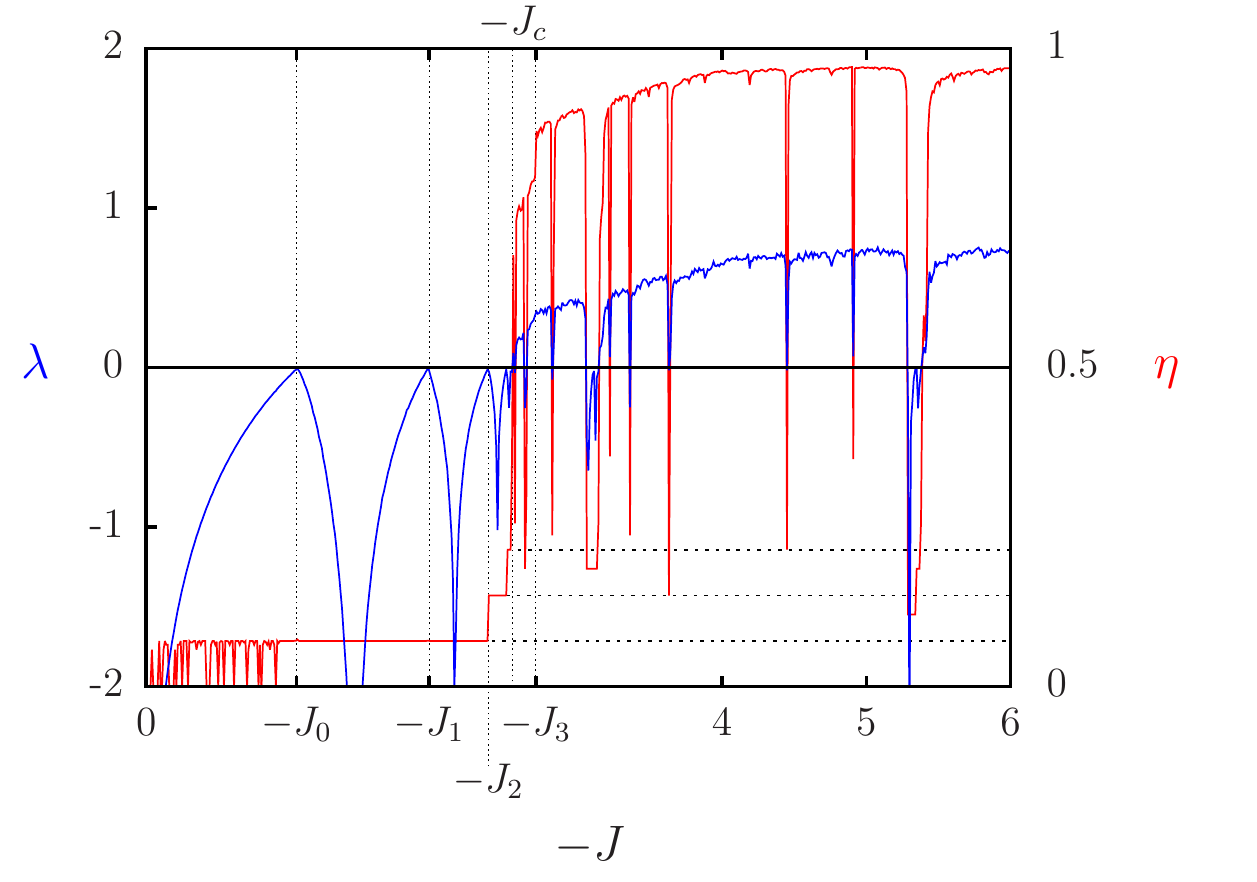}
 \end{center}
 \caption{\label{fig:mfb} (Color online)  (a) Bifurcation diagram of
   the mean field map, Eq.~\eqref{eq:mf}, by varying $J$.  The doubling bifurcation route to 
   chaos ends at $J=J_c$. For $0>J\geq J_2$ and $J_3>J\geq 6$ there is  
   only one attractor (blue dots). For $J_2>J\geq J_c$ there are
   two, one of them corresponds to the lower branches that bifurcate up to
   $J_c$ (red dots), and the other one to the upper branches (blue dots).  For $J_c>J\geq J_3$
   there are two chaotic attractors, one corresponding to the lower branches (blue dots), the other
   to  the top branches (red dots).
   For every value of $J$, the dots
   are 64 iterates of the map of Eq.~(\ref{eq:mf}) after a transient of
   $10^3$ time steps. For values of $J$ with only one basin of attraction
   the orbits do not depend on the initial average opinion $c(t=0)$.
   For values of $J$ that correspond to two attractors, one of them was
   found with $c(0)=0.1$, the other one with $c(0)=0.9$.
   (b) The Lyapunov exponent $\lambda$, top
   curve on the left of the graph (in blue), and entropy $\eta$, top
   curve on the right of the graph (in red), For every value of $J$,
   $\lambda$ was evaluated during $10^3$ time steps.  The entropy $\eta$is computed using
    $L=2^{14}$ bins. After a transient
   of 500 time steps, the probability distribution was evaluated during the next $100\times L$ time
   steps. The horizontal dotted lines are drawn, starting from below,
   at $\eta=w/m$, $w=1,\dots,3$
   corresponding to periodic orbits of period $2^w$.  The connectivity is  $k=20$ and
    the vertical dotted lines are drawn at $J_0=-1.045$, 
   $J_1=-1.965$, $J_2=-2.375$, $J_c=-2.545$, and $J_3=-2.705$. }
\end{figure}

\begin{figure}
 \begin{center}
  (a) \\
  \hskip -10mm
  \includegraphics[width=0.85\columnwidth]{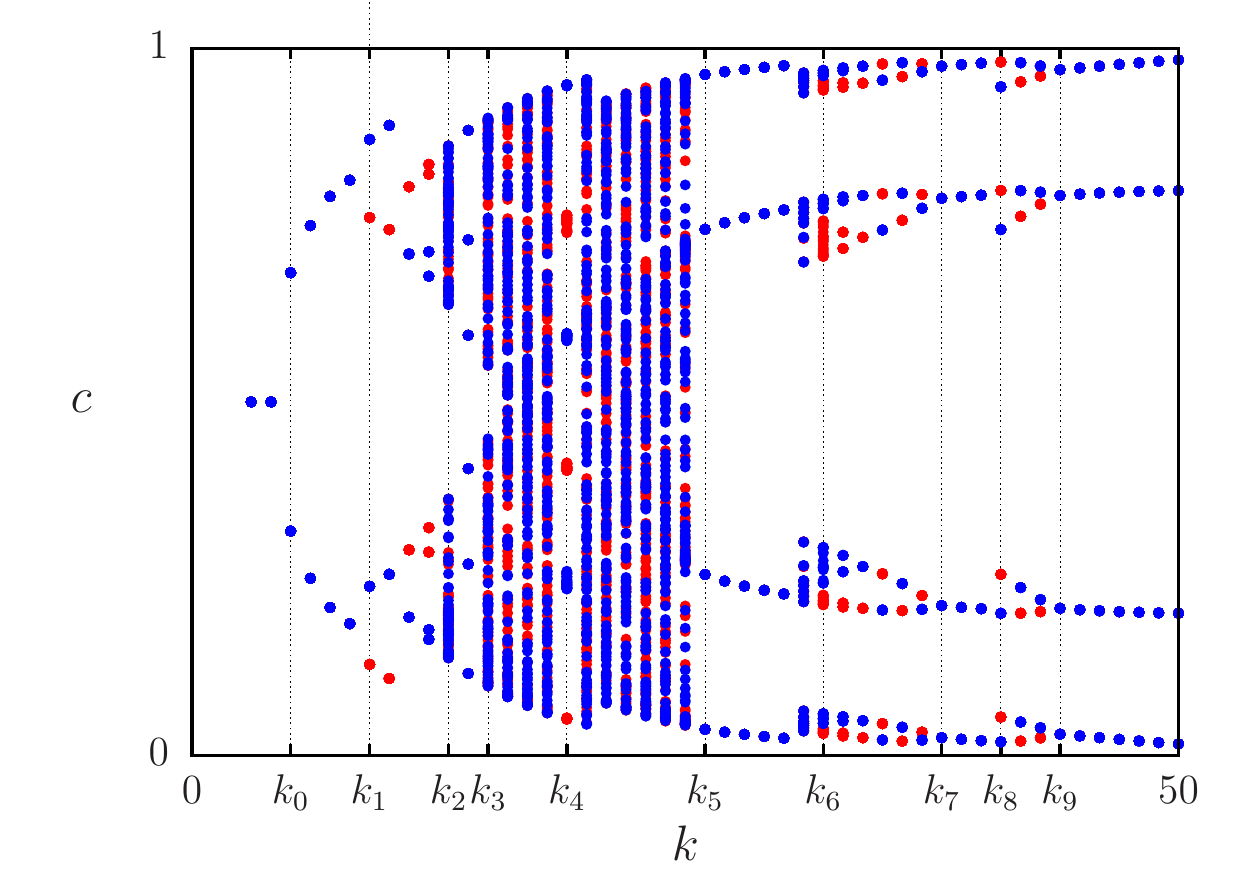}\\
  (b) \\
  \includegraphics[width=\columnwidth]{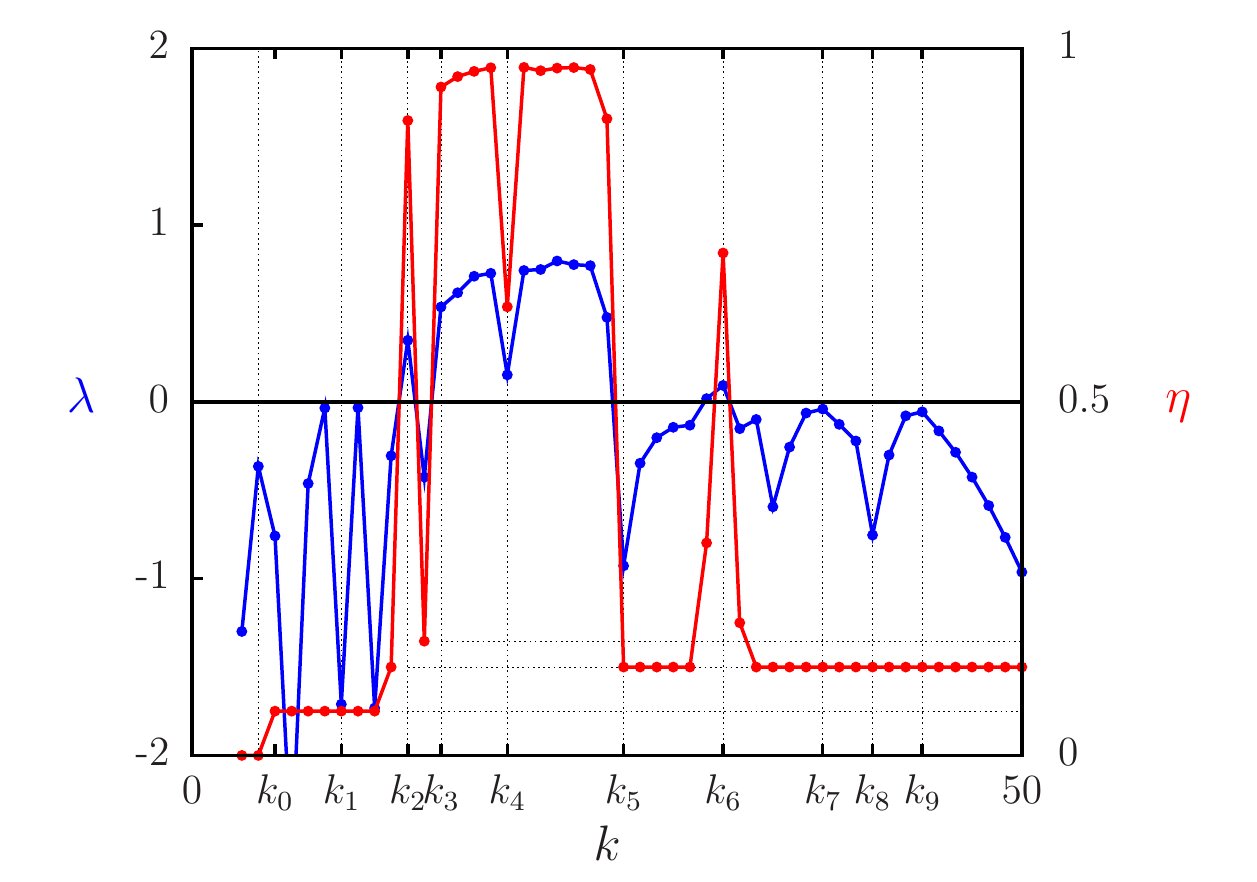}   
  \caption{\label{fig:mfb-k} (Color online.)  (a) Bifurcation diagram
    of the mean field map of Eq.~\eqref{eq:mf} varying $k$ for $J=-6$.
     For every value of $k$, two initial 
    values were considered, $c(0)=0.1$ and $c(0)=0.9$, and for each one 64
    iterations were plotted after a transient of $10^3$ time
    steps. For $k<k_0$ there is a fixed point and for $k_0\leq k<k_1$
    period-two orbits. For $k_1\leq k <k_2$ the bottom branches (in red)
    correspond to
    one attractor and the top branches (in blue) to the other one. 
    For $k_3\leq k<k_5$ the orbits are chaotic but for
    $k=k_4$ there are two attractors, one (in red) corresponds
    to the alternate clusters of points starting from below, the other one
    (in blue) to the other three clusters of points. For $k_6\leq k<k_7$, and
    $k_7\leq k<k_8$ there are again two attractors, one cluster (in red)
    corresponds roughly to the bottom branches and the other one (in blue) to
    the top branches. These attractors are not chaotic except for $k=k_6$. 
    (b) The Lyapunov exponent $\lambda$, top curve for $k<k_2$
    (in blue), and the entropy $\eta$, top curve for $k_3<k<k_5$ (in
    red), both as functions of the connectivity $k$ for the same values
    of $J$ as in (a). For each value of $k$,
    $\lambda$ was evaluated during $10^3$ time steps. For $\eta$,
    $L= 2^{16}$
    and the probability distribution was evaluated
    during the $100\times L$ time steps after a transient of $10^3$ time steps. 
    The horizontal dotted lines correspond, starting from below, to 
    perio-two, period-four, and period-six orbits. The vertical dotted lines are drawn at
    $k_0=5$, $k_1=9$, $k_2=13$, $k_3=15$, $k_4=19$, $k_5=26$, $k_6=32$,
    $k_7=38$, $k_8=41$, and $k_9=44$.}
 \end{center} 
\end{figure}

The mean-field map, Eq~\eqref{eq:mf}, shows a bifurcation
 diagram when varying the parameter $J$ (Figs.~\ref{fig:mfb} (a) 
 and \ref{fig:mfb-k} (a)). Since the mean-field map is deterministic, 
 these bifurcations can be characterized by means of the Lyapunov exponent $\lambda$. 
However, in order to study these diagrams and that produced by the stochastic microscopic simulations on the same ground, let us introduce the  Boltzmann entropy~\cite{Boltzmann} $\eta$ of the collective variable $c$. This entropy
is a good measure of disorder and in the case
of deterministic maps, large values of $\eta$ correspond to
positive values of the Lyapunov exponent as we show below. For probabilistic
processes, it is a measure of disorder. We define the normalized 
Boltzmann entropy $\eta$ as
\begin{equation}
 \label{eq:s}
 {\eta}=\dfrac{-1}{\log L}\sum_{i=1}^Lq_i\log q_i,
\end{equation}
where the interval $[0,1]$ is divided in $L$ disjoint intervals $I_i$ of equal size (bins) and
$q_i$ is the probability that  $c\in I_i$, $i=0,\dots,L-1$.  It is
clear that $0\leq \eta\leq 1$, the lower bound corresponding to a
fixed point, the upper one to the uniform distribution $q_i=1/L$. 
The probabilities $q_i$ are found numerically by finding the fraction
of time one orbit visits the bin $I_i$. 

The map $f$ of Eq.~(\ref{eq:mf}) depends on the parameters $J$, $k$, $q$ and
$\eps$. We keep $q$ and $\eps$ fixed.
By changing $J$ for $k=20$ we find the bifurcation diagram shown in
Fig.~\ref{fig:mfb} (a) with the corresponding values of the Lyapunov
exponent $\lambda$ and the entropy $\eta$ in Fig.~\ref{fig:mfb} (b).
The bifurcation diagram appears to show a period-doubling cascade, but it is
more complex than that.
For $0>J\geq J_0$ there are period-one orbits and for
$J_0>J\geq J_1$ period-two orbits. For $J_1>J\geq J_2$ the orbits
appear to have period four, but actually correspond to two separate period-two attractors. In other words, for $J=J_1$ there is a pitchfork bifurcation. For $J_2>J\geq J_c$ there are two separate period-doubling
bifurcations with the appearance of chaos at $J=J_c$. For $J_c>J\geq J_3$
there are two chaotic attractors that merge at $J=J_3$~\cite{bdr-col}. 
Due to the symmetry of the map, Eq.~(\ref{eq:sym}), if $c$ belongs to one of the basins of attraction, $1-c$ belongs to the other one.

In Fig.~\ref{fig:mfb-k} (a) we show the bifurcation diagram of the map $f$
as $k$ changes with fixed $J$, $\eps$, and $q$. For $k<k_0$ there are
period-one orbits and for $k_0\leq k< k_1$ period-two orbits. For 
$k_1\leq k <k_2$ there are two period-two attractors. The two attractors are
again present for $k=k_4$, $k_6\leq k<k_7$, and for $k_8\leq k<k_9$. 
For $k=k_4$ and $k=k_6$ the two attractors are chaotic. 
In Fig.~\ref{fig:mfb-k} (b) we show $\lambda$ and $\eta$ as
$k$ changes. Again, chaotic orbits have entropy larger than $\eta_c=1/2$.
Chaotic orbits are present for $k=k_2$, $k_3\leq k \leq k_5$, and $k=k_6$.
Both bifurcation diagrams, Figs.~\ref{fig:mfb} (a) and \ref{fig:mfb-k} (a)
are symmetric around $c=0.5$, a consequence of the symmetry of the
mean field map, Eq.~(\ref{eq:sym}).

\begin{figure}
 \begin{center}
  (a) \\
  \includegraphics[width=\columnwidth]{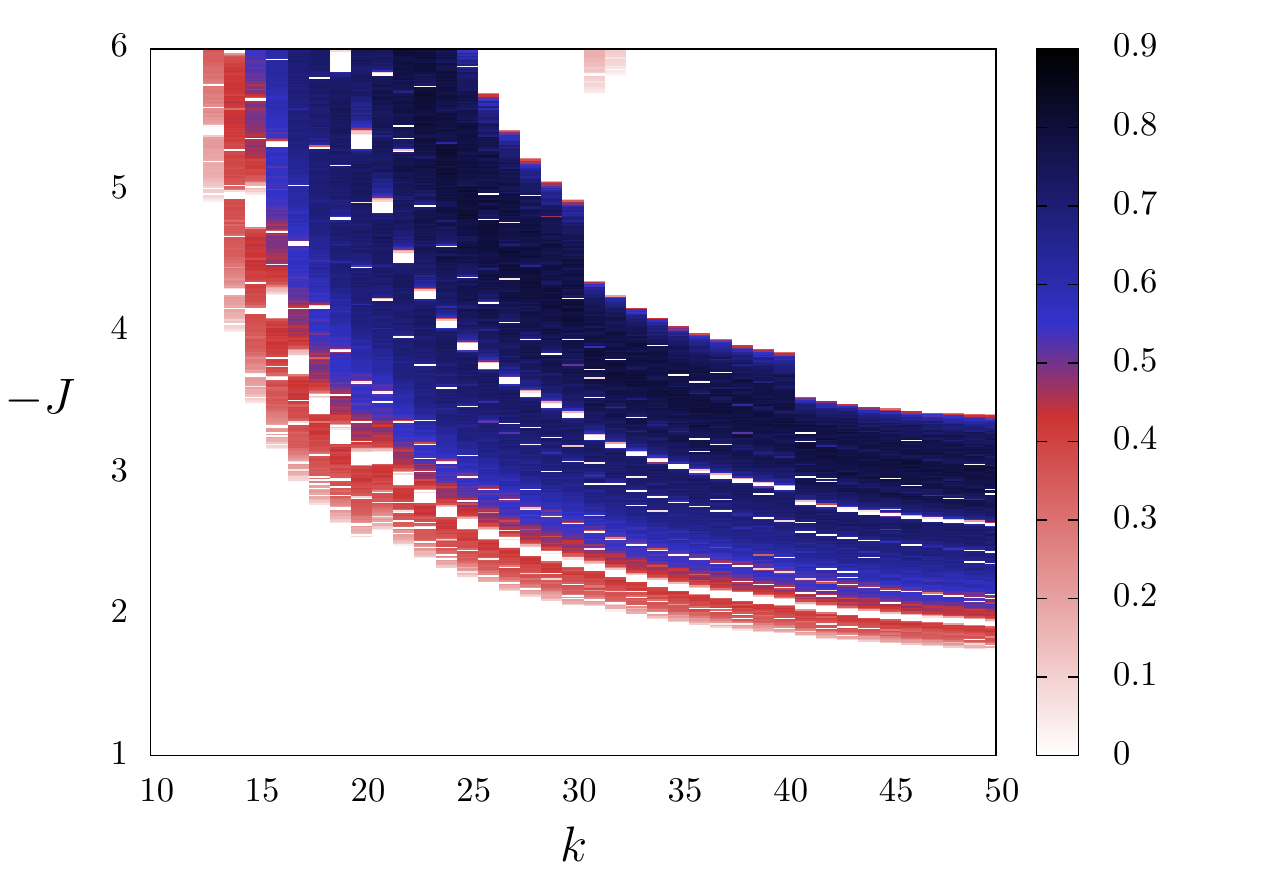}\\
  (b)\\
  \includegraphics[width=\columnwidth]{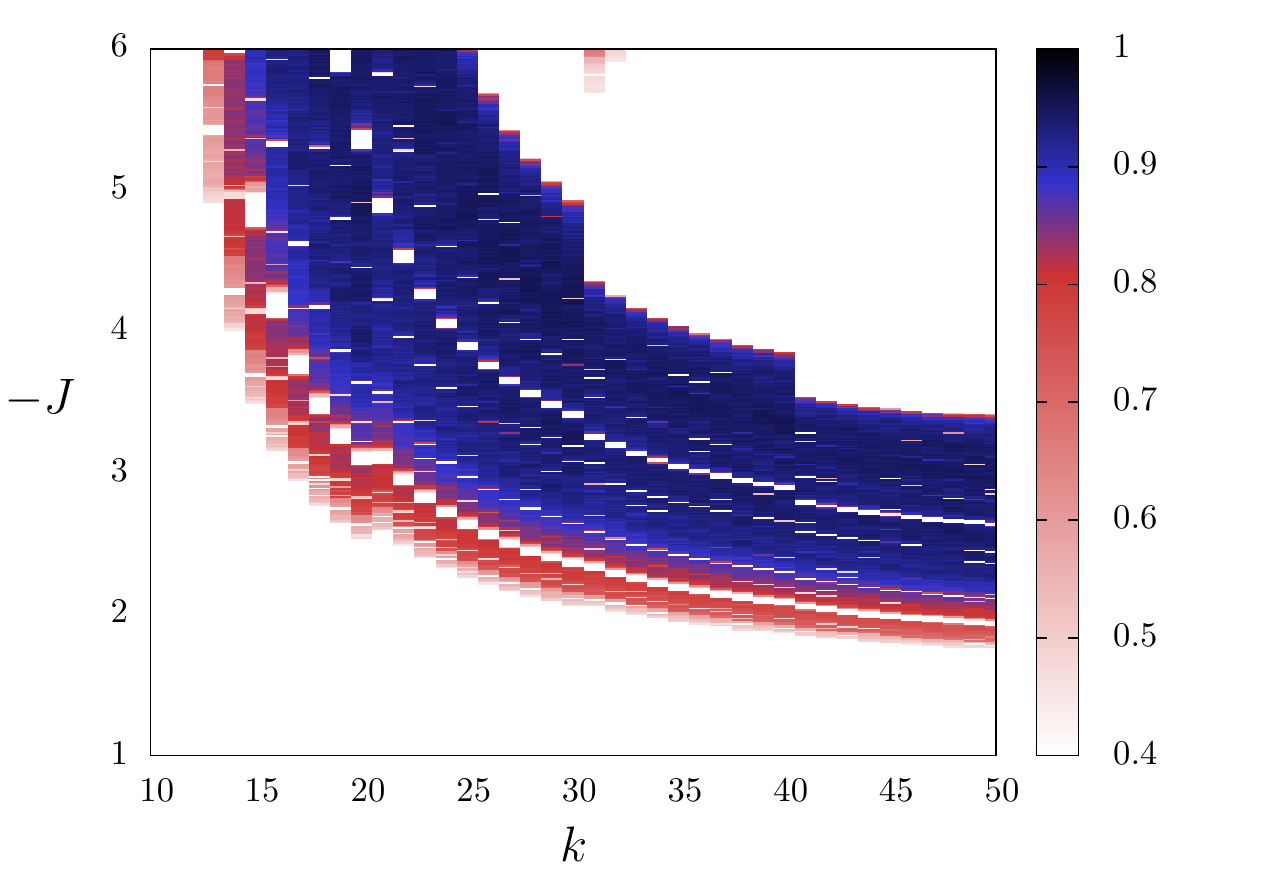}
  \caption{\label{fig:lyap-diag} (Color online) (a) Phase diagram
   showing positive values the Lyapunov exponent  $\lambda$
    for the mean field approximation Eq.~\eqref{eq:mf} as a function
    of $k$ and $J$. For each value of
    $J$ and $k$ the Lyapunov exponent $\lambda$ was calculated during
    $10^3$ time steps.  (b) Phase diagram of the entropy $\eta$
    showing values larger than 1/2, for the same values of $\eps$ and $q$
    as in (a). After a transient of $2\cdot10^3$ time  steps the probability distribution
     was evaluated during the next $10^3$ time steps on $L=128$ bins.
   }
 \end{center}
\end{figure}

In Figs.~\ref{fig:lyap-diag} (a) and (b) we show the phase diagrams, as
$J$ and $k$ change, of the Lyapunov exponent $\lambda$ and the entropy $\eta$
respectively. In (a) the points correspond to $\lambda>0$
and in (b) to $\eta>\eta_c=1/2$. These figures show that both quantities
are a good measure of chaos in this case. The values of $\lambda$ and $\eta$
shown in Fig.~\ref{fig:mfb} (b) correspond to those on the vertical line
$k=20$ of Fig.~\ref{fig:lyap-diag} (a) and (b) respectively. The results shown in 
Fig.~\ref{fig:mfb-k} (b) correspond to the horizontal lines
$J=-6$ of Fig.~\ref{fig:lyap-diag} (a) and (b).  

%


\section{Small-world networks}\label{sec:smallworld}

\begin{figure}
 \begin{center}
  \begin{tabular}{cc}
    (a) & (b)\\
    \includegraphics[width=0.45\columnwidth]{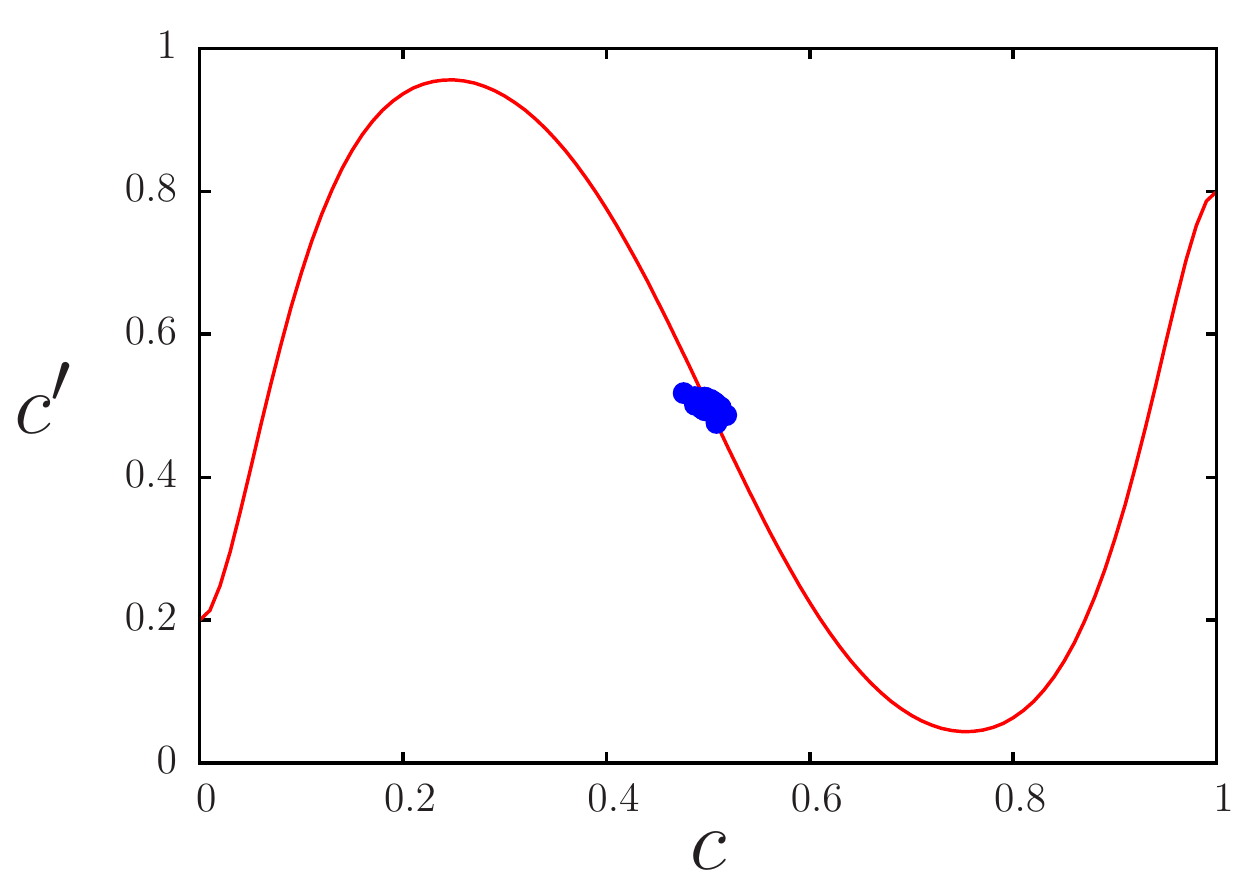}&
    \includegraphics[width=0.45\columnwidth]{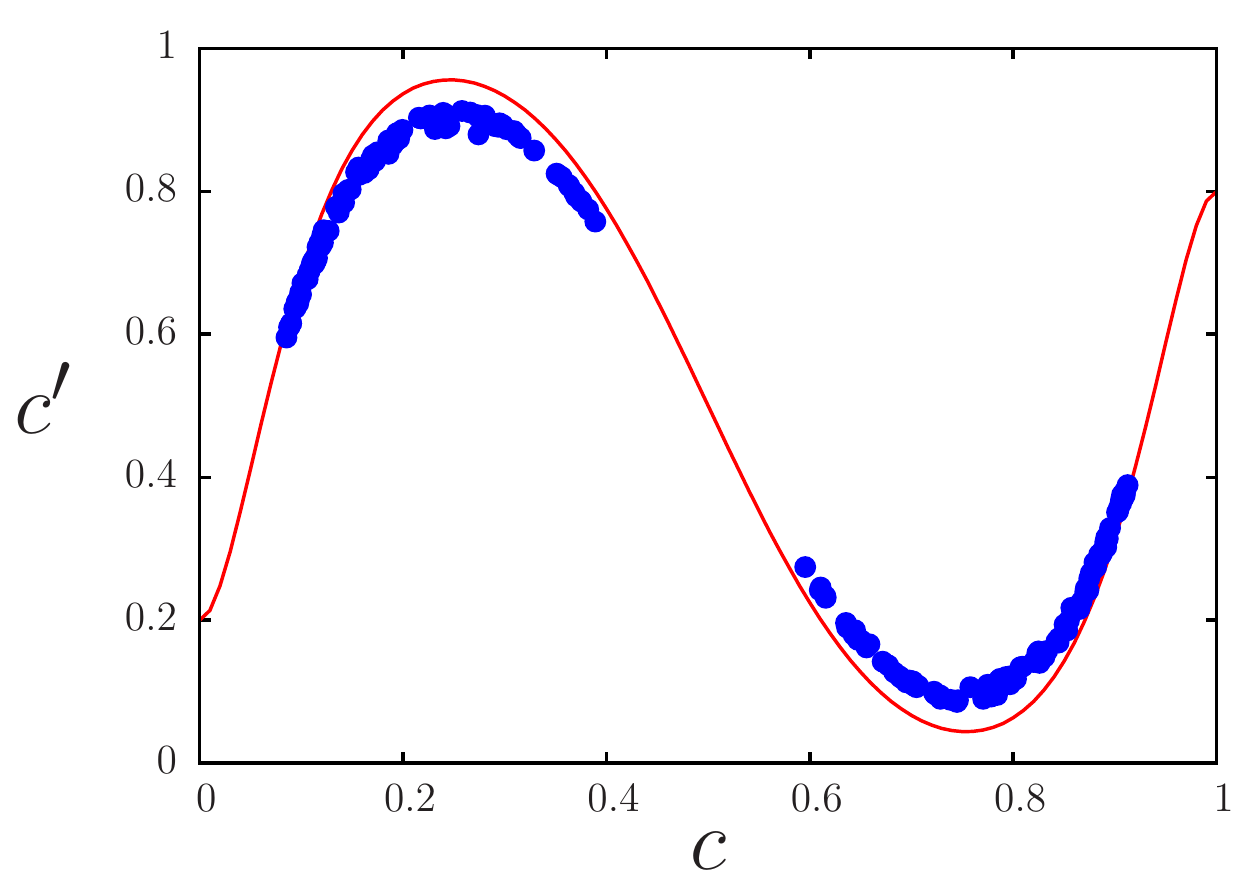} \\
    (c) & (d)\\
    \includegraphics[width=0.45\columnwidth]{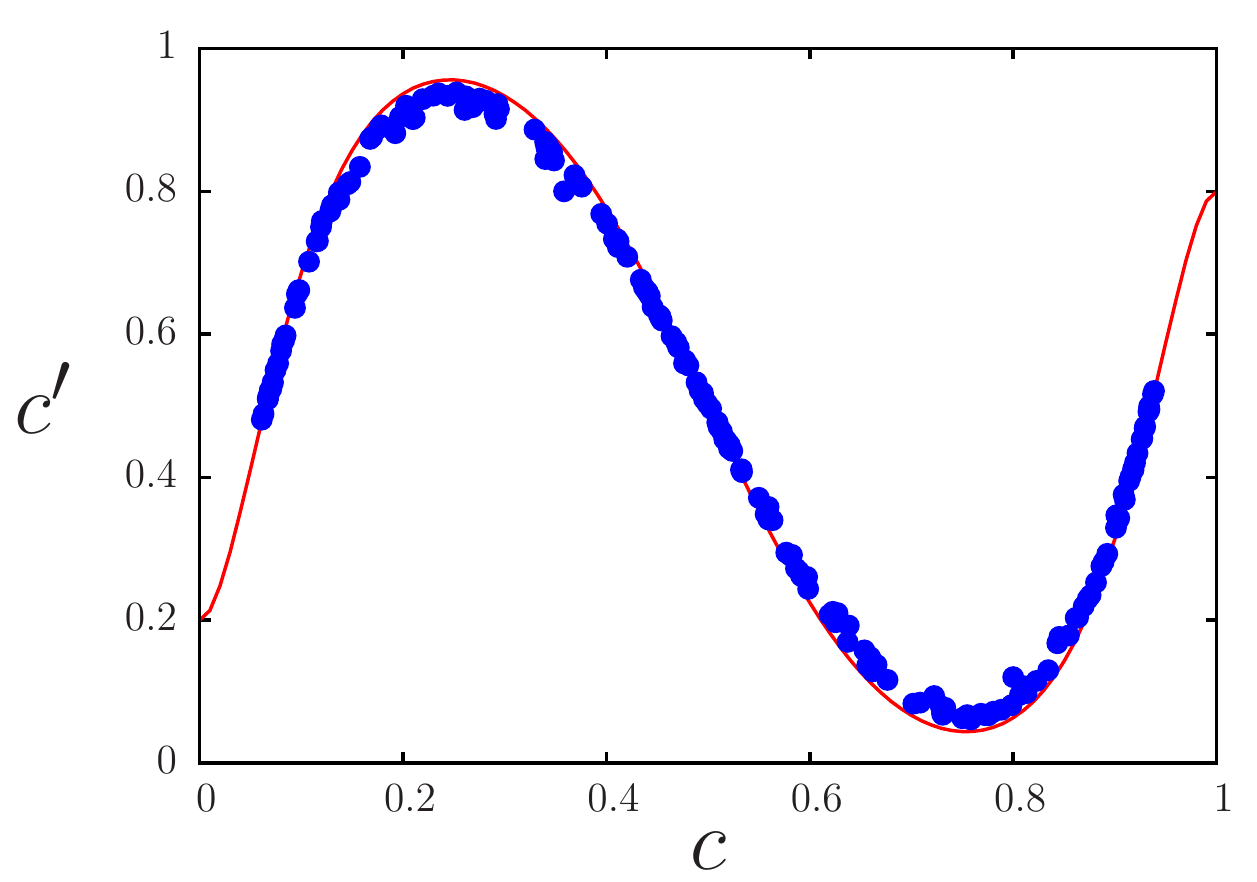} &
    \includegraphics[width=0.45\columnwidth]{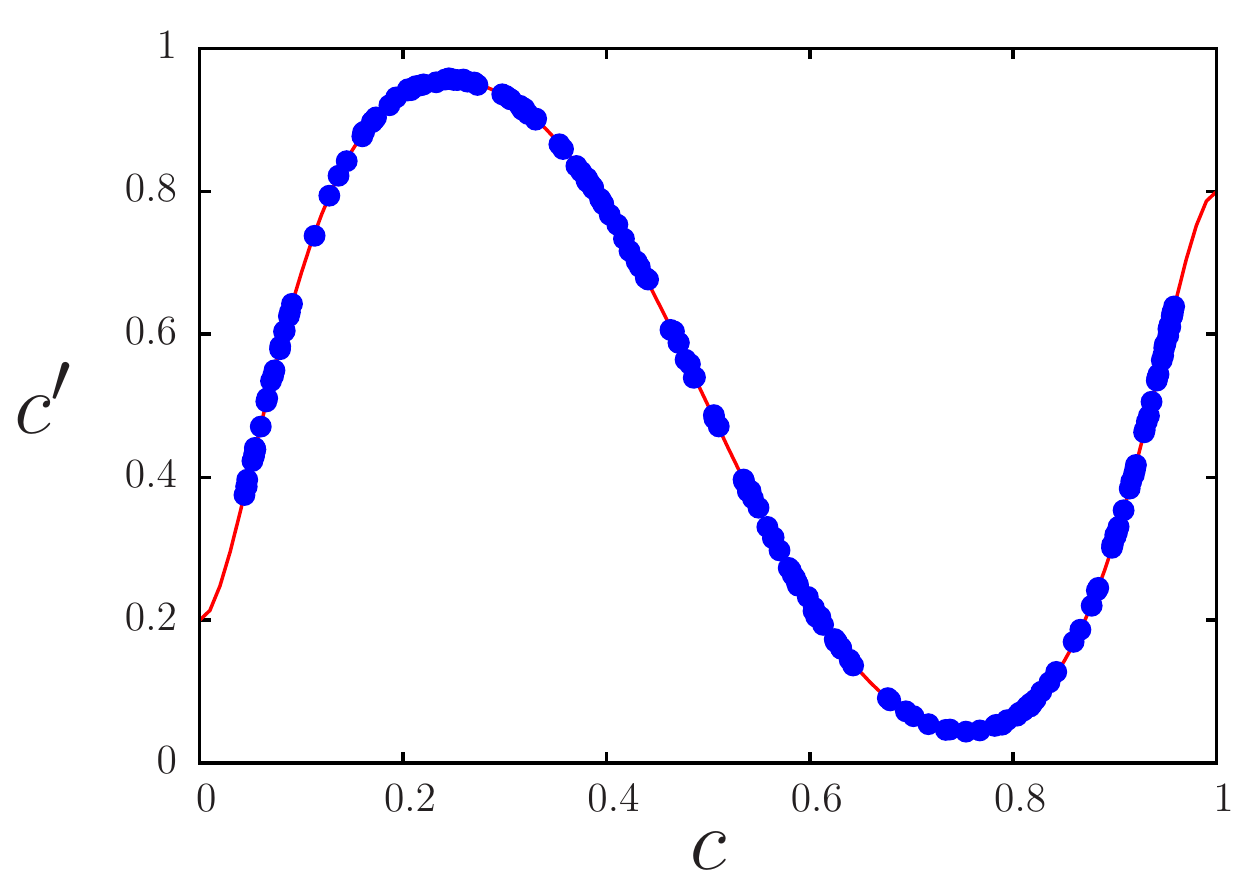} \\                                        
  \end{tabular}
  \end{center}
  \caption{\label{fig:conf} (Color online)  Return map of the
    average opinion $c$ on small-world networks for
    several values of the long-range connection probability $p$ with
    $J=-6$, $k=20$, $N=10^3$, and a transient
    of $10^3$ time steps. The following 200 iterations are shown as
    (blue) dots. The (red) continuous curve is Eq.~(\ref{eq:mf}). (a)
    $p=0.0$, (b) $p=0.5$, (c) $p=0.6$, and (d) $p=1.0$.}
\end{figure}

In the Watts-Strogatz small-world network model there is a smooth change from a regular to a random lattice~\cite{WattsStrogatz}. Starting with
a network with $N$ agents, where the neighborhood of each agent is formed by
his $k$ nearest neighbors, 
with probability $p$ each neighbor is replaced by another individual chosen at 
random.  
We call $p$ the long-range
connection probability. In Fig.~\ref{fig:conf} we show the return map
of the average opinion $c$, after a long transient, together with the
mean field return map $f$ of Eq.~(\ref{eq:mf}) for several values of
$p$. For $p=0$, the density $c$ fluctuates around its mean value
0.5. As $p$ grows, the system becomes more homogeneous and the
distribution of points approaches the mean field behavior, even though
the mean field approximation has been derived by imposing the absence
of correlations. As shown in the figure, for $p=0.6$ the return map is
already close to the mean field behavior and for $p=1$ it is
indistinguishable from it.

\begin{figure}
 \begin{center}
  \begin{tabular}{cc}
   (a) & (b) \\
   \includegraphics[width=0.45\columnwidth]{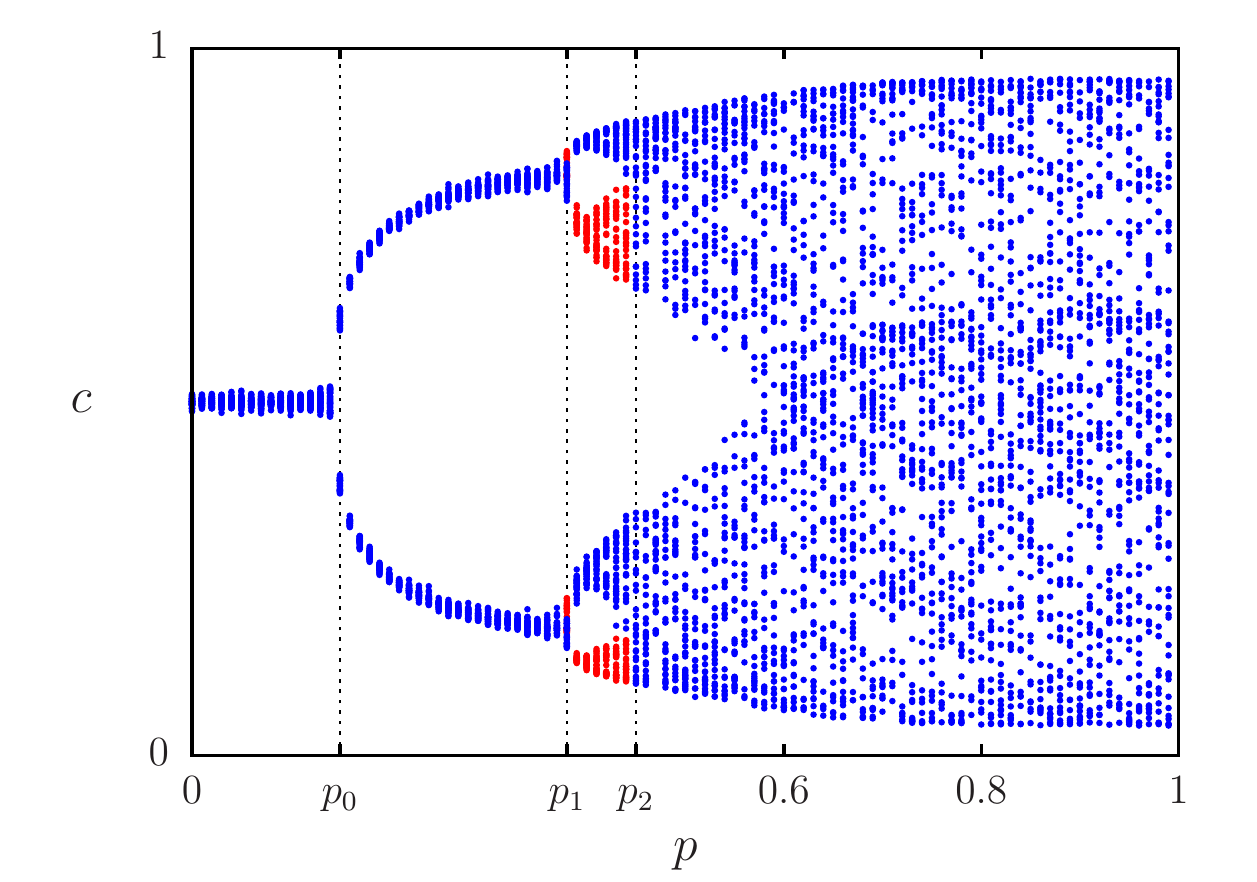} &
   \includegraphics[width=0.45\columnwidth]{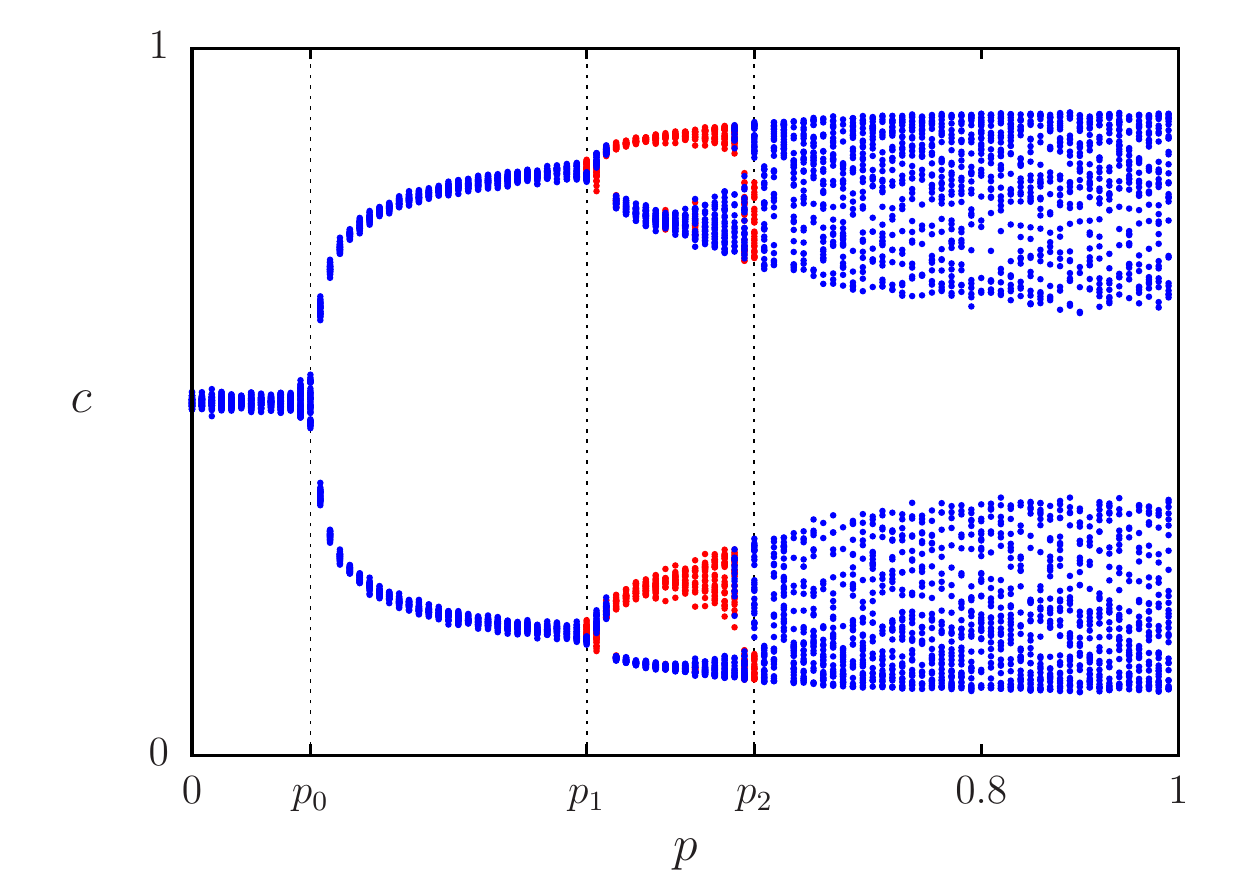} \\
   (c) & (d) \\
   \includegraphics[width=0.45\columnwidth]{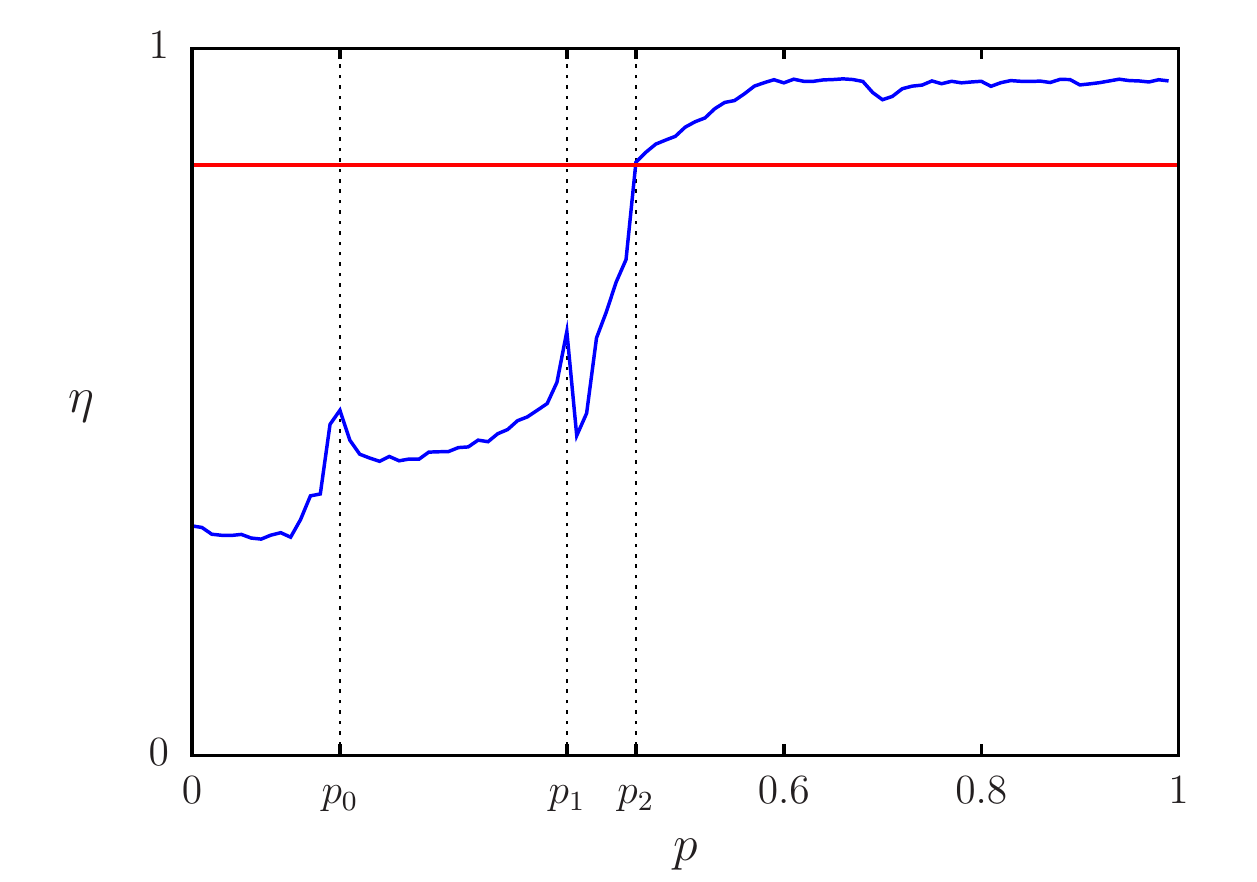} & 
   \includegraphics[width=0.45\columnwidth]{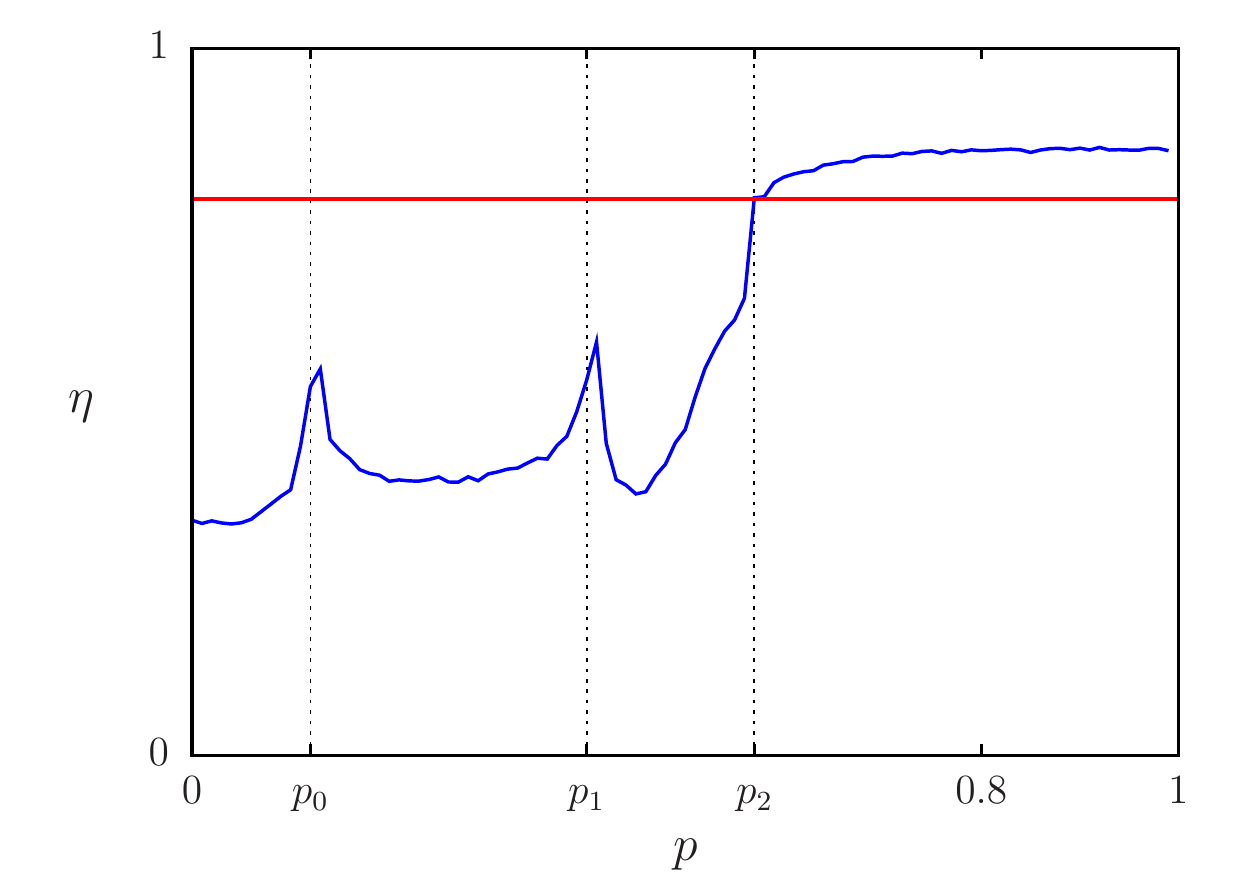} 
  \end{tabular}
 \end{center}
  \caption{\label{fig:swbp} (Color online.)  (a) and (b) Small-world
    probabilistic bifurcation diagrams as functions of the long range
    probability $p$. For $p\lesssim p_0$ there are almost periodic orbits of period one
    and for $p_0\lesssim p\lesssim p_1$ of period two. For $p_1\lesssim p\lesssim p_2$
    we find two attractors, one (in red) in the lower branches,
    the other one (in blue) in the top ones. (c) and (d) The entropy $\eta$
    as a function of $p$. The (red) lines mark the value of $\eta_d$.
    (a) and (c) $J=-6$, $p_0\sim 0.15$, 
    $p_1\sim 0.38$, $p_2\sim 0.45$, and $\eta_d=\eta(p_2)=0.835$. 
    (b) and (d) $J=-3$, $p_0\sim 0.12$, $p_1\sim 0.40$, 
    $p_2\sim 0.57$, and $\eta_d=\eta(p_2)=0.787$. In (a) and (b)
    the number of agents is $N=5\cdot 10^4$, the connectivity is $k=20$.
     After a transient of $4\cdot 10^3$ time steps, the probability distribution
    is evaluated using $L=256$ bins during the next 
    $100\times L$  time steps.}
\end{figure}

We show in Figs.~\ref{fig:swbp} (a) and (b) the probabilistic bifurcation
diagrams of $c$ as a function of the probability of long-range connections $p$
for  $J=-6$ and $J=-3$ and the same value of $k$.  In both figures, 
for $0<p\lesssim p_0$ and $p_0<p\lesssim p_1$ we can identify  period-one
and period-two orbits respectively. For $p_1\lesssim p\lesssim p_2$
there are two period-two attractors which become indistinguishable for
$p\simeq p_2$. For $p_2\lesssim p$ there is only one attractor. 
In  Figs.~\ref{fig:swbp} (b) and (c) we show the
corresponding entropy. We would like to find a threshold $\eta_d$ for
the appearance of disorder, similar to $\eta_c$ of the mean field
approximation, and  we propose $\eta_d=\eta(p_2)$ shown as the horizontal
lines in Figs.~\ref{fig:swbp}.

\begin{figure}
 \begin{center}
  \begin{tabular}{cc}
   (a) & (b) \\
   \includegraphics[width=0.45\columnwidth]{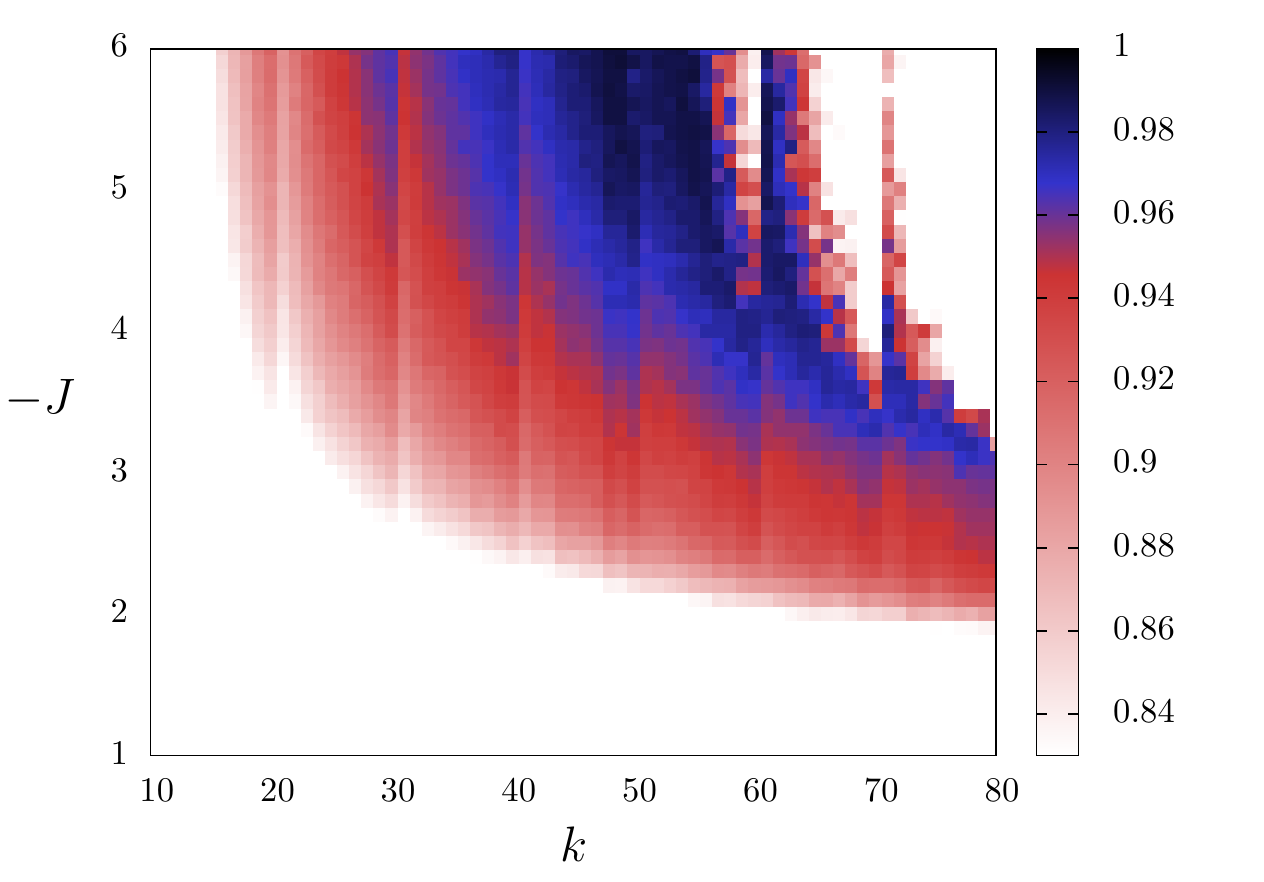} &
   \includegraphics[width=0.45\columnwidth]{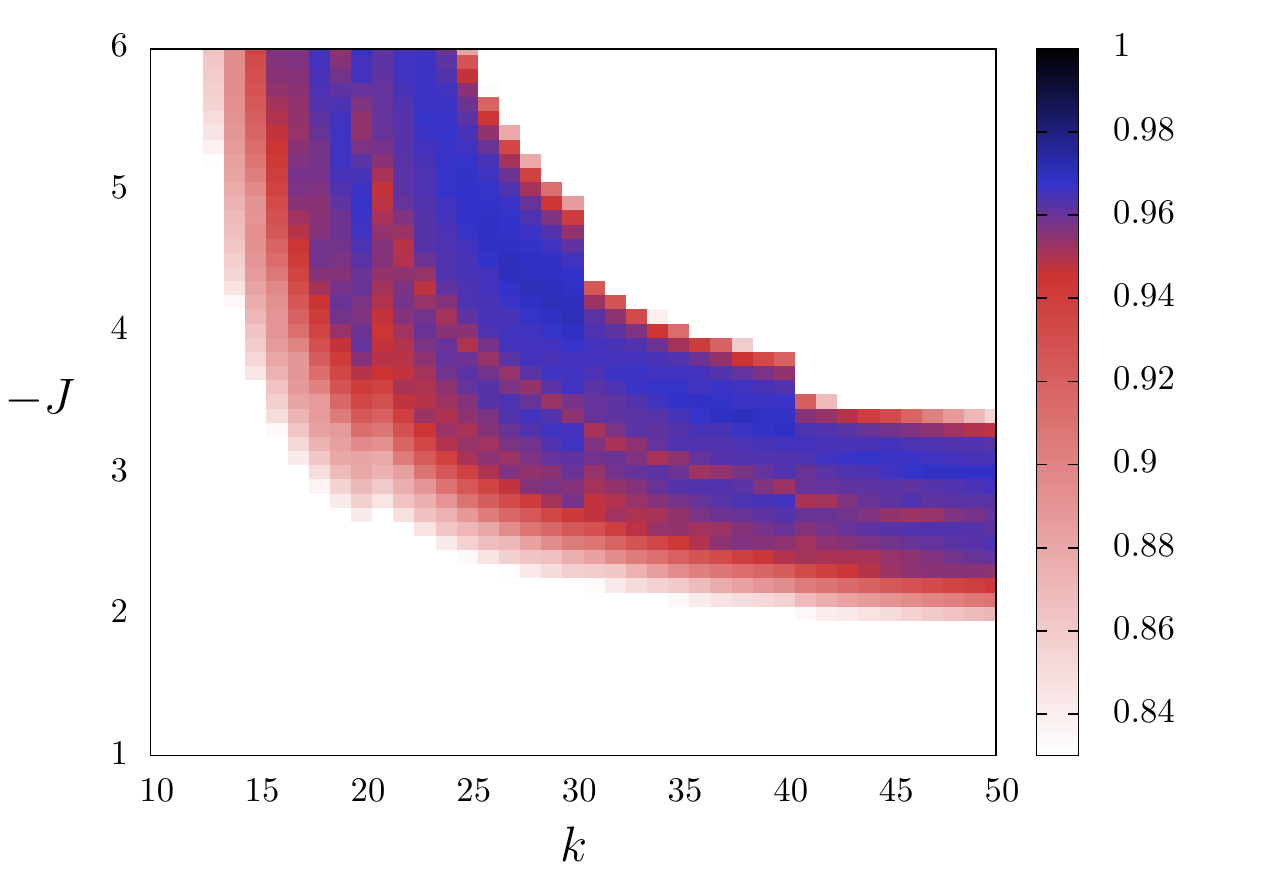} 
  \end{tabular}
 \end{center}
 \caption{\label{fig:swbif} (Color online) Entropy phase diagrams 
 of simulations on small-world networks as functions of $k$ and $J$ for
 (a) $p=0.5$ and (b) $p=1.0$. The colored region corresponds to 
 $\eta>\eta_d=0.8$.  Entropy computed with 128 bins, 
 lattice size $N=10^4$, sampling time $1.2\cdot10^4$ steps after a transient of $4\cdot 10^3$ steps.}
\end{figure}

In Fig.~\ref{fig:swbif} (a) and (b) we show the phase diagrams of the entropy $\eta$ 
for $p=0.5$ and $p=1$ respectively.  It is evident that for $p=1$, Fig.~\ref{fig:swbif} (a), the diagram is very similar to that of Fig.~\ref{fig:lyap-diag}, while for $p=0.5$, Fig.~\ref{fig:swbif} (a), there is a sort of dilatation of the high-entropy region, extending to larger values of $k$ (and beginning also with higher values of $k$. The dependence on $J$ is much less marked. It is possible to roughly understand these results assuming  that the main contributions to the  mean-field character of the collective behavior come from the fraction of links that are rewired (long-range connections), that depends on $p$.

\section{Scale-free networks}
\label{sec:networks}

Human and technological networks often present a scale-free character, with different degrees of correlation among nodes. 
In this section we present results of the model on uncorrelated scale-free 
networks~\cite{Barabasi}. Starting from a fully connected group of
$m$ agents, other $N-m$ agents join sequentially, each one choosing $m$
neighbors among those already in the group. The choice is
preferential, the probability that a new member chooses agent $i$ is
proportional to its connectivity $k_i$, the number of neighbors agent
$i$ already has. Another way of building the network is choosing a random 
edge of a random node and connecting to the other end of the edge, since such
an edge arrives to a vertex with probability proportional to 
$kp(k)$~\cite{newman-PRE-64-026118-2001}.
\begin{figure}
 \begin{tabular}{cc}
  (a)  & (b) \\
  \includegraphics[width=0.45\columnwidth]{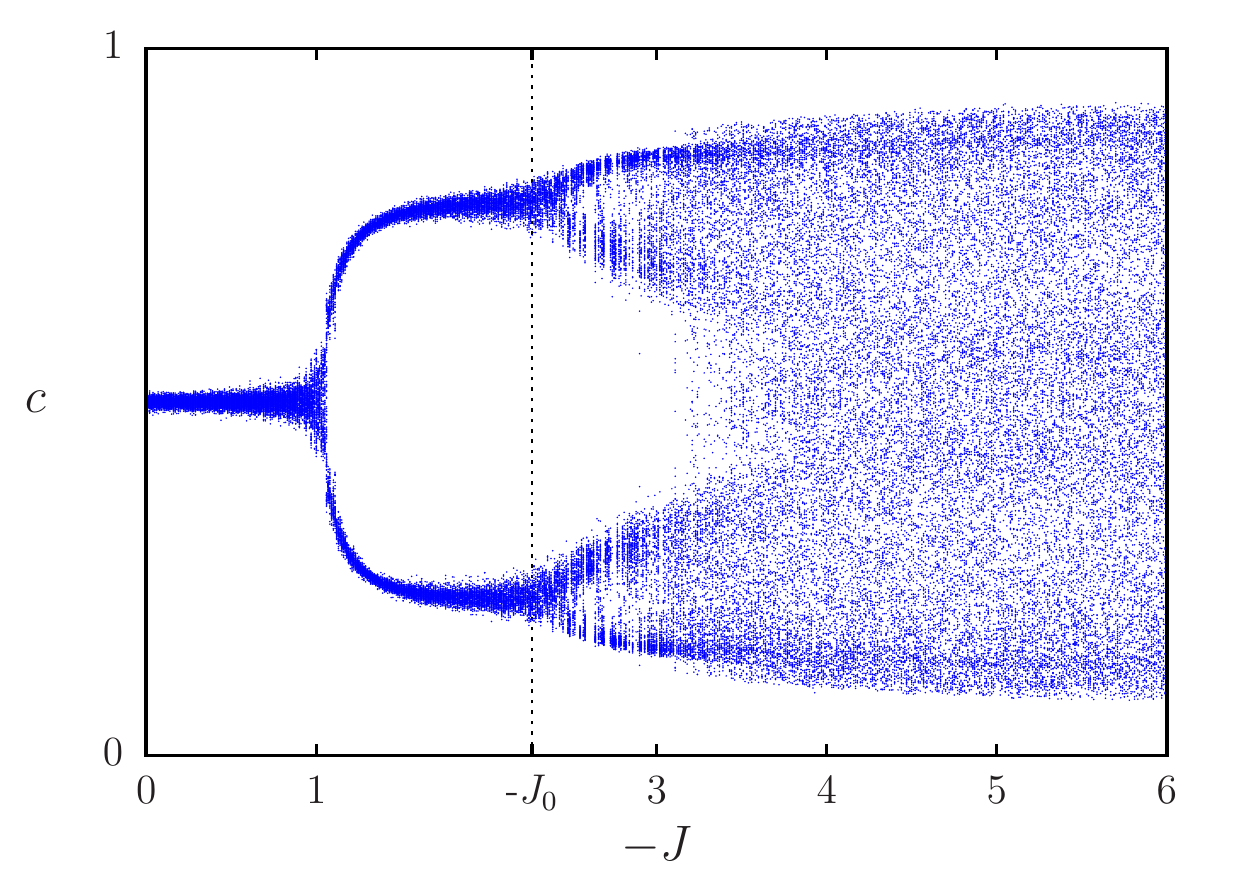} &
  \includegraphics[width=0.45\columnwidth]{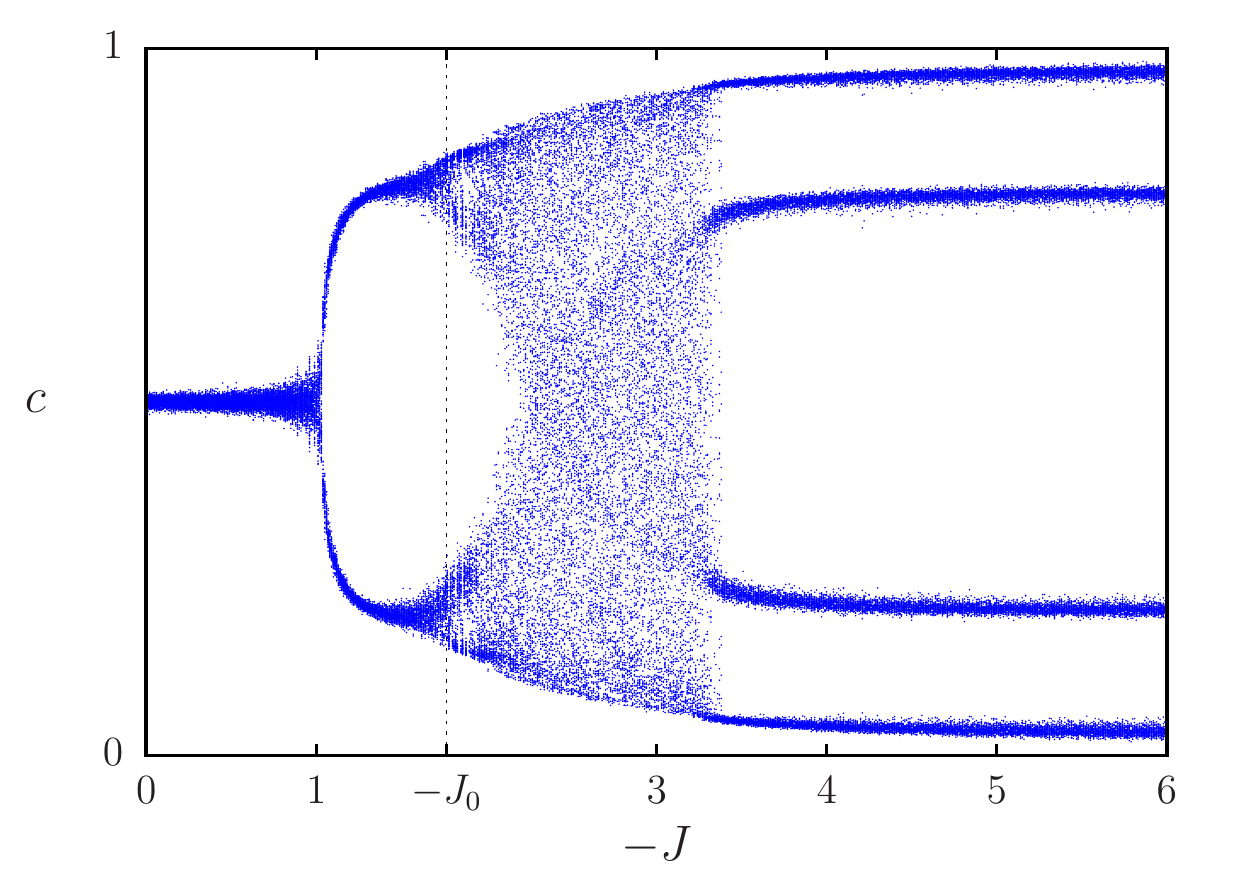}\\ 
  (c) & (d) \\
  \includegraphics[width=0.45\columnwidth]{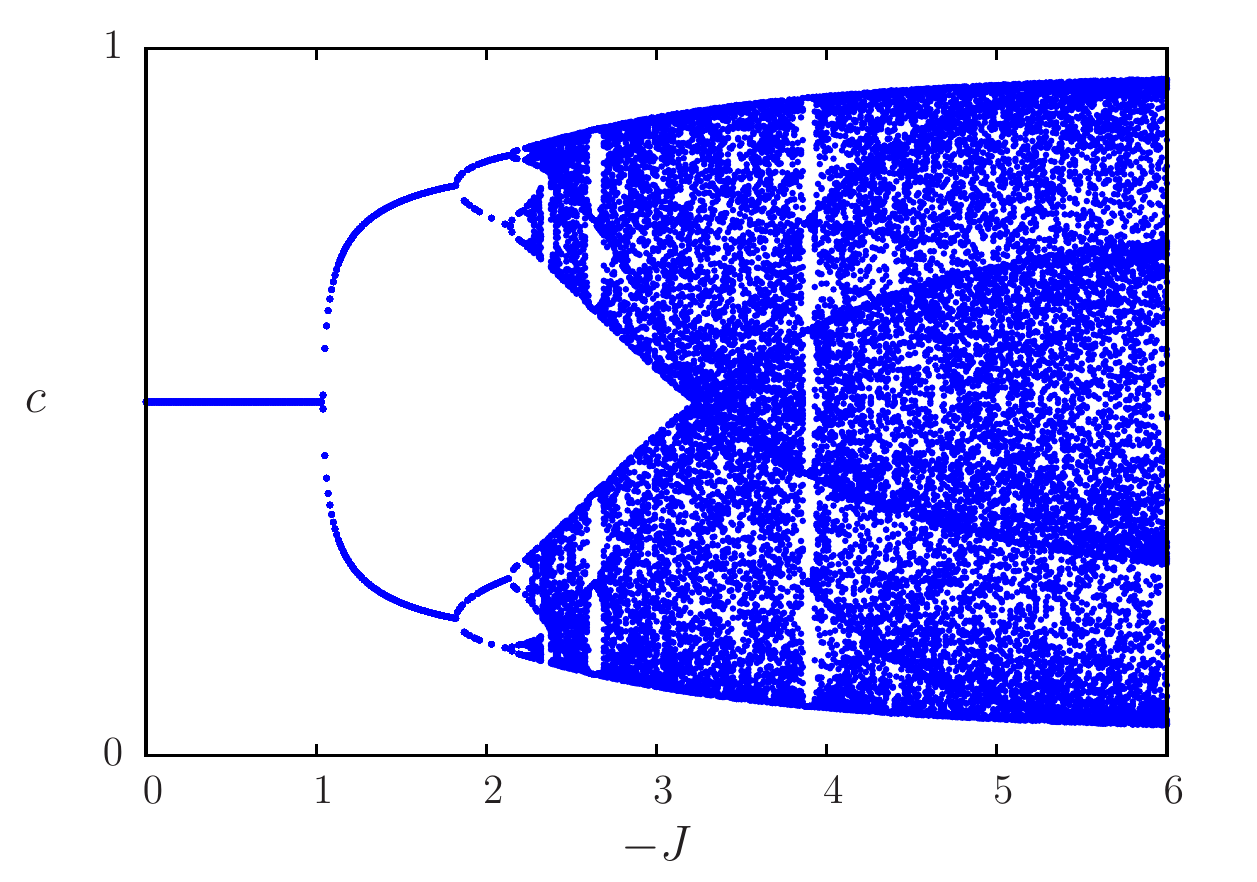}& 
  \includegraphics[width=0.45\columnwidth]{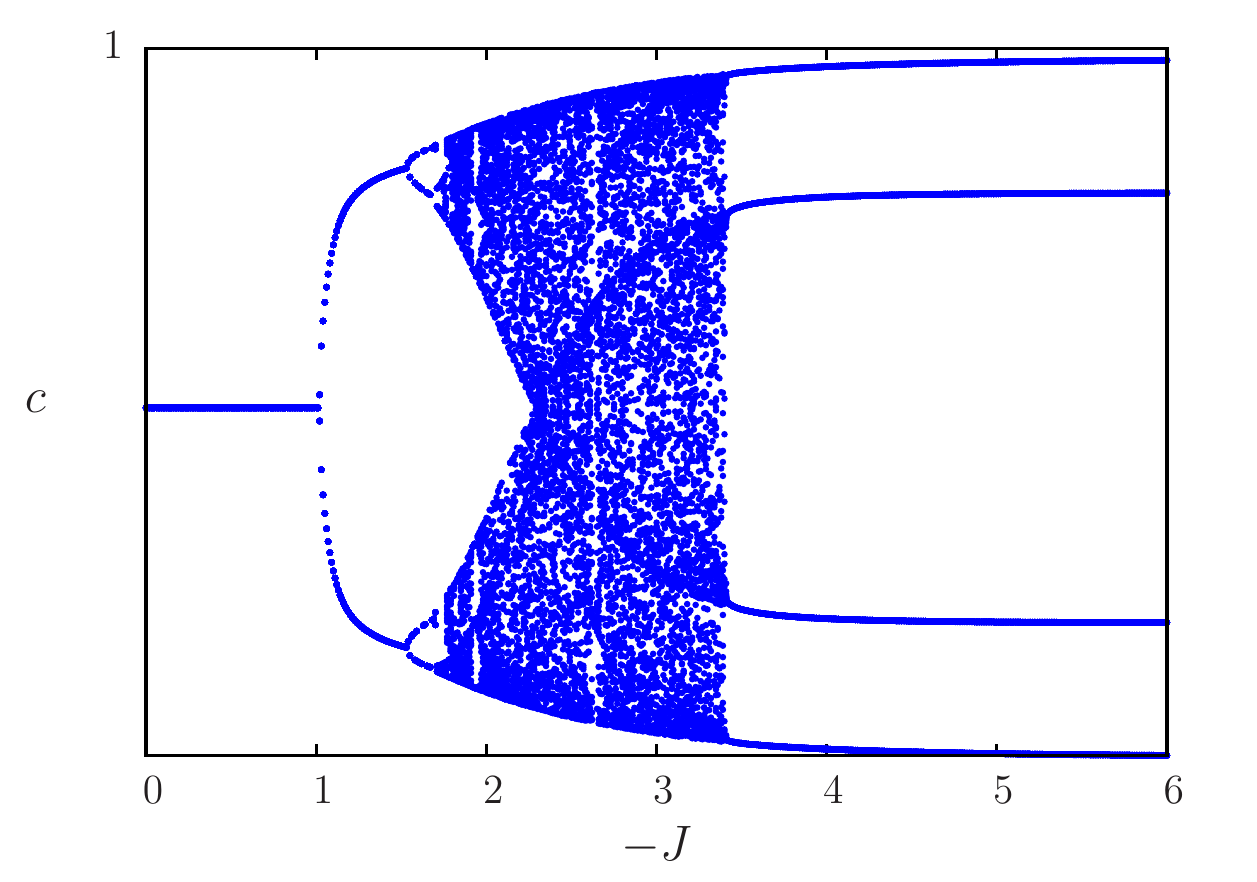}\\
  (e) & (f) \\
  \includegraphics[width=0.45\columnwidth]{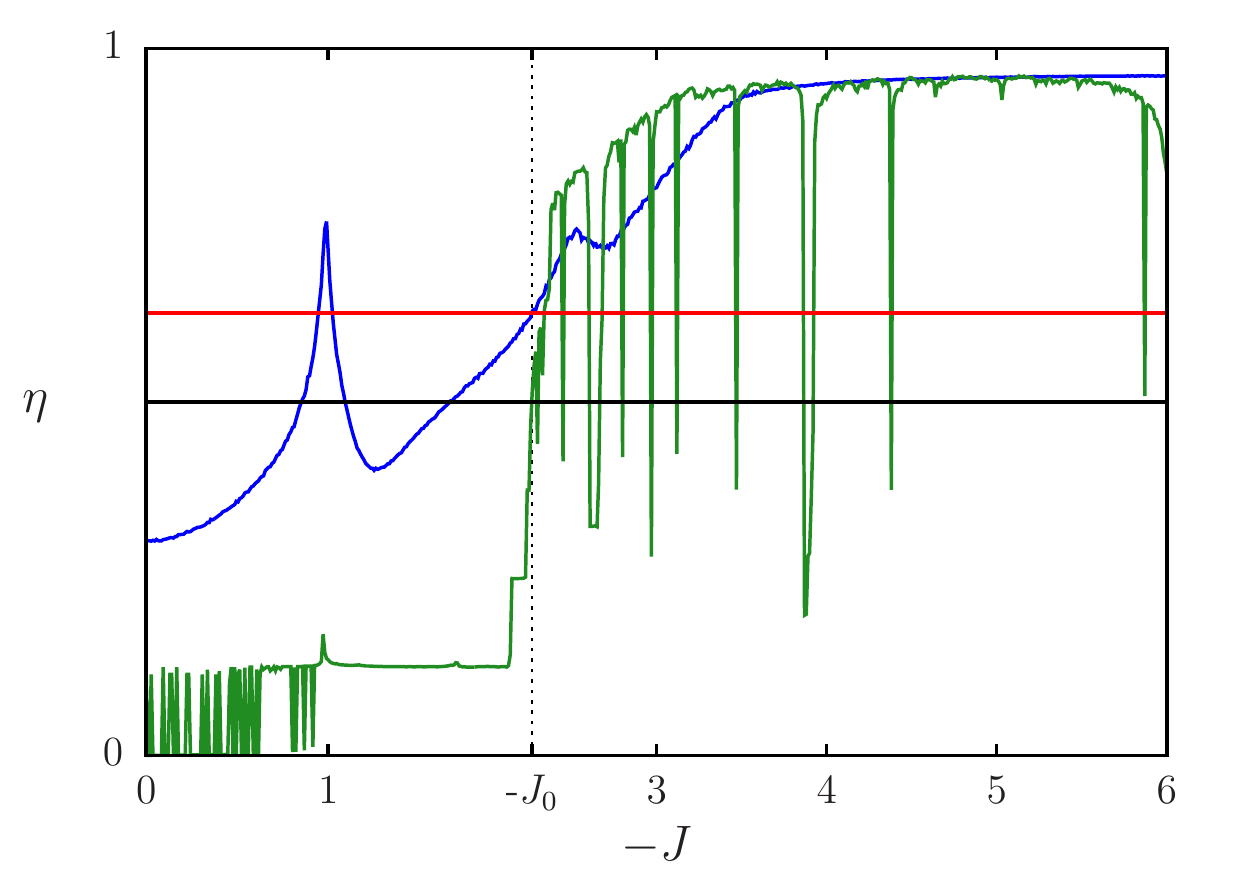} &
  \includegraphics[width=0.45\columnwidth]{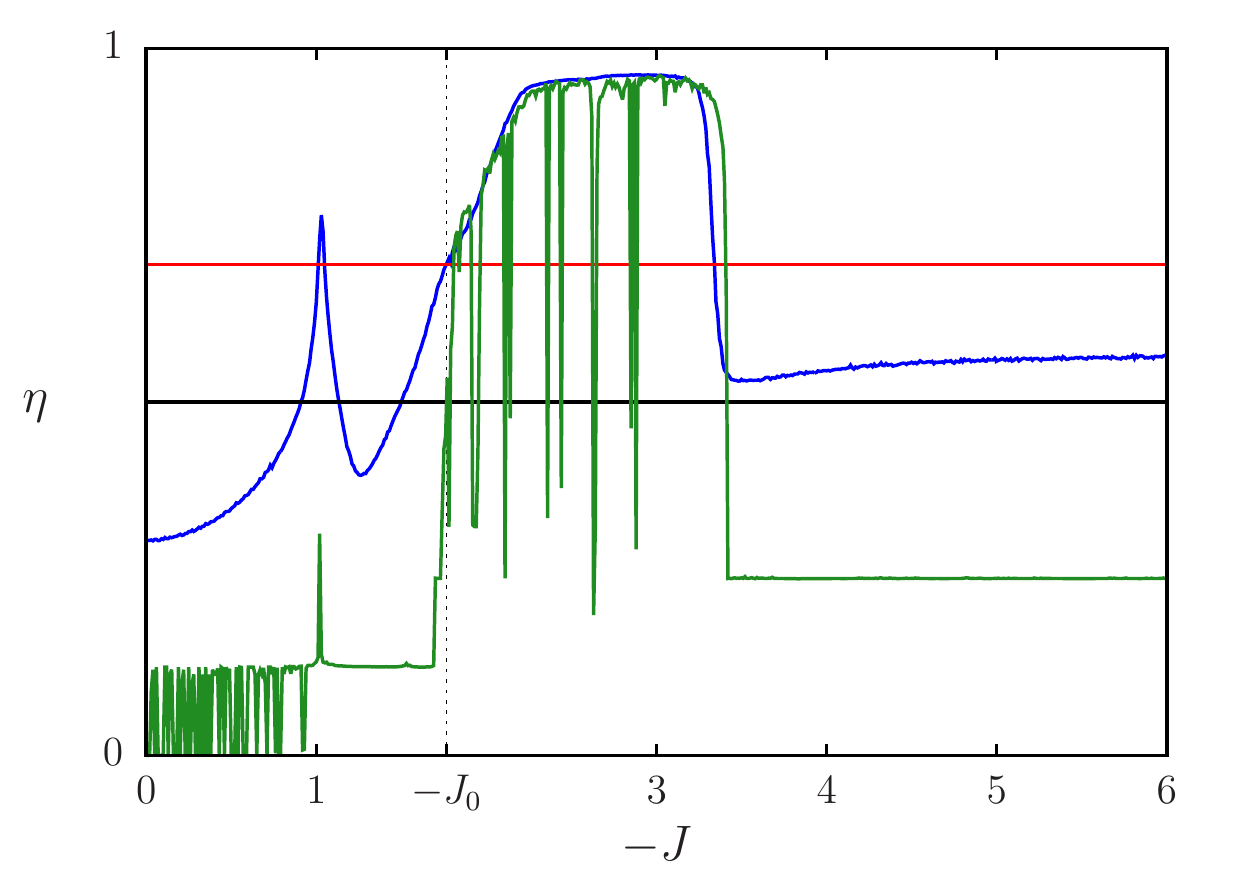}
 \end{tabular}
 \caption{\label{fig:sfn-bif} (Color online) Comparisons between simulations on scale-free networks and 
 mean-field approximation. In (a) and (b) bifurcation 
   diagrams of simulations on scale-free networks with $m=15$ 
   and $m=30$, respectively. In both cases, the number of agents is $N=10,000$ and
   for every value of $J$, the initial opinions are chosen at random with
   $c=1/2$, and 128 values of $c$ are plotted after a transient of 300 time
   steps. In (c) and (d) we show the bifurcation diagrams of 
   the mean field approximation with $k=25\sim 1.7\times 15$ and $k=51\sim 1.7\times 30$, respectively.
   In both figures, $c(0)$ is chosen at random and 
   128 values of $c$ are plotted after a transient of $10^3$ time steps. In (e) and (f) the entropy $\eta$ of the 
   simulations on scale-free networks (top curve -- blue
   for $J>-1$) is compared with that of the mean-field approximation, 
   (bottom curve -- green for $J>-1$). In (e), $k=15$ and $m=25$, in (f) $k=30$ and $m=51$.
   The entropy is found by dividing the unit interval
   in 256 bins. For each value of $J$, $c(0)=0.1$ and after
   a transient of $10^3$ time steps the entropy was evaluated during
   the following $2.56\cdot10^4$ time steps. In (a) and (e) $J_0=-2.267$,
   and in (b) and (f) $J_0=-1.766$.} 
\end{figure}

In the Appendix we show that
the model dynamics on scale-free networks is comparable
to the mean field approximation of Sec.~\ref{sec:meanfield} on a network
with constant connectivity $k$ with 
\begin{equation}
 \label{eq:km}
 k=\alpha m,\qquad \alpha\sim 1.7.
\end{equation}
In Figs.~\ref{fig:sfn-bif} (a) and (b) we show the probabilistic 
bifurcation diagrams of the
model on scale-free networks as a function of $J$ for two values of $m$
and in Figs.~\ref{fig:sfn-bif} (c) and (d) we show the
bifurcation diagram of the mean field approximation, Eq.~(\ref{eq:mf}), for the corresponding
values of $k$ according to Eq.~(\ref{eq:km}). We find a qualitative 
agreement between these bifurcation diagrams.

In Figs.~\ref{fig:sfn-bif} (e) and (f) we show the entropy of
of the mean field approximation and of the simulations on 
scale-free networks. We find a reasonable agreement when $\eta>\eta_d$ with
$\eta_d=\eta(J_0)$ with $J_0$ the value of $J$ for which the entropy of the mean field approximation
crosses the line $\eta=1/2$ for the first time. Thus, the entropy is a good way of comparing
both dynamics when $k$ and $m$ are related according to Eq.~(\ref{eq:km}). Above 
$\eta_d$, both entropies are numerically similar,
except where there are periodic windows in the mean field approximation,
and this agreement is better for $m=30$ and $k=51$.


\section{Conclusions}
\label{sec:conclusions}

We studied a reasonable contrarian opinion model. The reasonableness
condition forbids the presence of absorbing states. In the model, this
condition depends on two parameters that are held fixed. The model
also depends on the connectivity $k$ which may vary among agents, and
the coupling parameter $J$. The neighborhood of each agent is  defined
by an adjacency matrix that can have fixed or variable connectivity
(fixed or power-law) and a regular or stochastic character.

The interesting observable is the average opinion $c$ at time $t$.  We
computed the entropy $\eta$ of the stationary distribution of $c$,
after a transient.

In the simplest case, the neighborhood of each agent includes $k$ random sites. In this case,  the mean field approximation for the time evolution of the
average opinion exhibits,  by changing  $J$, a period doubling bifurcation
cascade towards chaos with an interspaced pitchfork bifurcation. A positive (negative) Lyapunov exponent corresponds to an entropy
larger (smaller) than $\eta_c=1/2$. Thus, entropy is a good measure of chaos
for this map, and can be also used in the simulations of  the stochastic microscopic model. 

The bifurcation diagram of the mean field approximation as a function of $k$
shows periodic and chaotic regions, also with a pitchfork bifurcation. Again,
entropies larger than $\eta_c$ correspond to chaotic orbits.

Actual simulations on a one-dimensional lattice show incoherent local oscillations  around $c=1/2$.
By rewiring at random a fraction $p$ of local connections, the model presents a series of  bifurcations induced by the small-world effect: the density $c$ exhibits a probabilistic bifurcation diagram that resembles that obtained by varying $J$ in the mean field approximation.
These small-world induced bifurcations are consistent with the general trend, long-range connections induce mean field behavior. This is the first observation of this for a system exhibiting a \emph{chaotic} mean field behavior. Indeed, the small-world effect makes the system more coherent (with varying degree). We think that this observation may be useful since many theoretical studies of population behavior have been based on mean field assumptions (differential equations), while actually one should rather consider agents, and therefore spatially-extended, microscopic simulations. The well-stirred assumption is often not sustainable from the experimental point of view. However, it may well be that there is a small fraction of long-range interactions (or jumps), that might justify the small-world effect. 

The  model on scale-free networks with a minimum
connectivity $m$ shows a similar behavior to that of the mean field 
approximation of the model on a network with constant connectivity $k$,
Eq.~(\ref{eq:mf}) if $k=\alpha m$ with $\alpha\sim 1.7$. 

In summary, we have found that, as usual, long-range rewiring leads to mean-field behavior, which can become chaotic by varying the coupling or the connectivity. Similar scenarios are found in actual microscopic simulations, also by varying the long-range connectivity, and in scale free networks. 

This study can have applications to the investigation of  collective phenomena in algorithmic trading.

\section*{acknowledgements} 
Interesting discussions with Jorge Carneiro and Ricardo Lima are
acknowledged. This work was partially supported by Recognition Project UE
grant n$^\circ$ 257756 and project PAPIIT-DGAPA-UNAM IN109213. 

\section*{Appendix}

The similarity between the bifurcations diagrams in Figs.~\ref{fig:mfb} (a)  and \ref{fig:mfb-k} (a), 
which comes out from the similarities of the mean-field maps when changing $J$ and $k$ (Fig.~\ref{fig:mfeq}),
 can be explained by using a continuous approximation for the connectivity $k$.  By using Stirling's approximation for the binomial coefficients in Eq.~\eqref{eq:mf}, for intermediate values of $c$~\cite{ott02}, we obtain
\begin{equation}\label{approx}
	\binom{k}{w} {c}^{w}(1-{c})^{k-w} \simeq \frac{1}{\sqrt{2\pi k c(1-c)}} \exp\left[\frac{-k\left(w/k-c\right)^2}{2 c (1-c)}\right].
\end{equation}
In this approximation, Eq.~\eqref{eq:mf}  can be written as
\begin{equation}\label{eq:mfapprox}
c'=\bigintss_{-\infty}^{\infty} \mathrm{d}x\,\sqrt{\frac{k}{2\pi c (1-c)}}\exp\!\!\left[-\frac{k (x-c)^2}{2c(1-c)}\right]\!\tau(x)
\end{equation}
with $x$ the continuous approximation of $w/k$. This expression is just a Gaussian convolution of  $\tau$,
 \textit{i.e.}, a smoothing of the transition probability, 
as can be seen by comparing Fig.~\ref{fig:tau-mf} with Fig.~\ref{fig:mfeq}. This
smoothing has the effect of reducing the slope of the curve in a way
similar to changing $J$ (but is depends also on $c$), and this explains the similarities
between the bifurcation diagrams in Fig.~\ref{fig:mfb} (a) and
\ref{fig:mfb-k} (a). For instance, Fig.~\ref{fig:mfb} (a) is obtained
for $k=20$, a value that in  Fig.~\ref{fig:mfb-k} (a) corresponds to a chaotic strip just
after a window with six branches. A similar window can be observed also in Fig.~\ref{fig:mfb} (a) by increasing $J$ from the value $J=-6$ of Fig.~\ref{fig:mfb-k} (a).

This approximation can be used also to find the ``effective'' connectivity of the model on a scale-free network. 
The mean field approximation for a non-homogeneous network can be written as
\begin{equation}
\label{eq:mfsf}
 c'_k = \sum_{\substack{s_1,s_2,\dots,s_k \\j_1,j_2,\dots,j_k}} 
    \prod_{i=1}^k c_{j_i}^{s_i}(1-c_{j_i})^{1-s_i} Q(j_i|k)%
    \tau(h_i),
\end{equation}
with $c'_k$ the probability that the opinion of an agent with connectivity
$k$ at time $t+1$ is one, and $c_j$ the probability that the opinion of an agent with
connectivity $j$ at time $t$ is one. 
The sum on the \textit{r.h.s} is taken over the opinions $s_1,\dots,s_k$
of the $k$ agents in the neighborhood, and over their connectivities
$j_1,\dots,j_k$. The
variables $s_i$ take the values zero or one, while $j_i$ ranges from
$m$ to $\infty$. The quantity $Q(j|k)$ is the probability that the
agent with connectivity $j$ is connected to another one of
connectivity $k$ and $\sum_j Q(j|k)=1$.

\begin{figure}
\centerline{\includegraphics[width=\columnwidth]{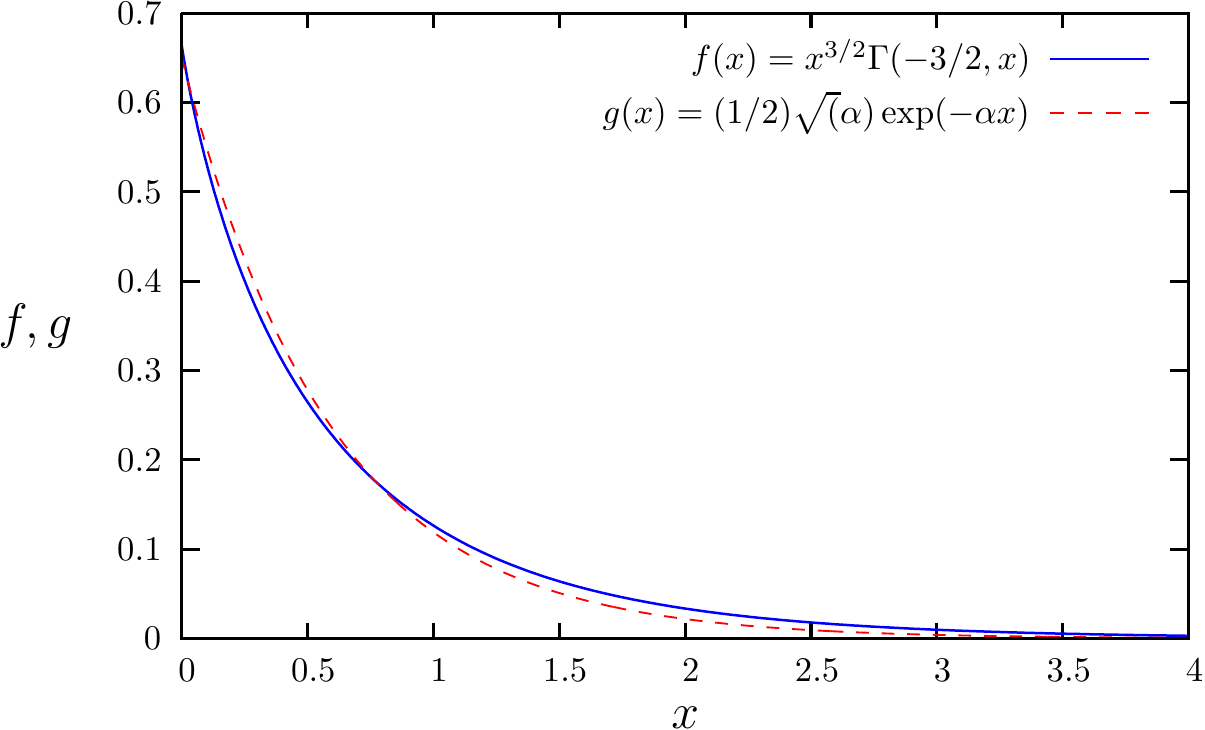}}
\caption{\label{fig:incomplete} (Color online) Comparisons between the two  functions $f(x)$ ((blue) solid line) and $g(x)$ ((red) dashed line), for $\alpha=1.7$, corresponding to the minimum of $\int_0^4 (f(x)-g(x))^2 \mathrm{d}x$ }
\end{figure}

Since this network is symmetric,  $kQ(j|k)P(k) = j Q(k|j)P(j)$ (detailed balance). 
It is also non-assortative, so $Q(j|k)$ does not depend on $k$ and we can write
$Q(j|k)=\phi(j)$. By summing the detailed balance condition over $j$ we get $\phi(k) = k P(k)/\langle k\rangle$. 
Therefore, Eq.~\eqref{eq:mfsf} becomes
\begin{equation}
  \label{eq:mfsf1}
  c'_k = \sum_{s_1,s_2,\dots, s_k} \tau\left(\frac{\sum_i s_i}{k}\right)\prod_{i=1}^k  \sum_{j_i} \dfrac{j_i P(j_i)}{\langle k\rangle} c_{j_i}^{s_i}(1-c_{j_i})^{1-s_i}.
\end{equation}

In the previous equation, $s_i$ is either zero or one, so that only
one between $c_{j_i}^{s_i}$ and $(1-c_{j_i})^{1-s_i}$ is different
from zero. Assuming that $c_k$ depends only slightly on $k$ in
Eq.~\eqref{eq:mfsf1}, we approximate $(\sum_{j_i} j_i P(j_i)
c_{j_i})/\langle k\rangle$ with $c$ and we get
\[
\begin{split}
c'_k &= \sum_{s_1,s_2,\dots,s_k}\tau\left(\frac{\sum_i s_i}{k}\right)\prod_{i=1}^k {c}^{s_i}(1-{c})^{1-s_i},\\
 &= \sum_w \tau\left(\frac{w}{k}\right)\binom{k}{w} {c}^{w}(1-{c})^{k-w} ,
 \end{split}
\]
with $w=\sum_i s_i$. 
In order to close the equation we average $c'_k$ over the probability distribution $P(k)$. 

By using the approximation of Eq.~\eqref{approx},we get
\[
	\begin{split}
	c'=&\sum_{k=m}^{\infty} P(k) \sum_w \tau\left(\frac{w}{k}\right)\binom{k}{w} {c}^{w}(1-{c})^{k-w} \\
	 \simeq& \quad \int_m^{\infty} \mathrm{d}k\; P(k) \\
	 &\qquad \int_{-\infty}^{\infty} \mathrm{d}x\; \tau\left(x\right)\sqrt{\frac{k}{2\pi c(1-c)}}  \exp\left[-\frac{k\left(x-c\right)^2}{2 c (1-c)}\right] ,
\end{split}	 
\]
where $x=w/k$. For scale-free networks the connectivity distribution $P$
is given by $P(k)=2m^2k^{-3}$. Then a
\[
\begin{split}
{c}' &\simeq \int_{-\infty}^{\infty} \mathrm{d}x \frac{2m^2\tau(x)}{\sqrt{(2\pi {c}(1-{c}))}} 
\int_m^\infty \mathrm{d} k\; k^{-5/2}\exp(-kA), \\
& = \int_{-\infty}^{\infty} \mathrm{d}x \frac{m^{1/2}\tau(x)}{(2\pi {c}(1-{c}))^{1/2}} 
2(mA)^{3/2}\Gamma\left(-\frac{3}{2}, mA\right), 
\end{split}
\]
where $A=A(x) = (x-c)^2/(2c(1-c))$ and $\Gamma(a,x)$ is the incomplete upper gamma function extended to negative values of $a$ (the function $x^{-a} \Gamma(a,x)$ is single-valued and analytic for all values of $a$ and $x$~\cite{Abra}).

The function $f(y)=y^{3/2}\Gamma(-3/2,y)$ is well approximated by  $g(y)=(1/2)\sqrt{\alpha} \exp(-\alpha y)$, as shown in Fig.~\ref{fig:incomplete}. Therefore we can write
\[
{c}' \simeq \int_{-\infty}^{\infty} \mathrm{d}x\;\tau(x) \sqrt{\frac{\alpha m}{2\pi {c}(1-{c})}}
\exp\left[-\frac{\alpha m(x-{c})^2}{2{c}(1-{c})}\right].
\]
This last expression has the form of Eq.~\eqref{eq:mfapprox}, with an effective connectivity $\tilde{k}=\alpha m$. 

Since the argument $y$ of $g(y)$ is $mA(x)=m(x-c)^2/(2c(1-c))$, the substituted $g(x)$  results to be a Gaussian, centered around $x=c$. The important values of  $g(x)$  lie between 0 and 4, depending on the value of $c$. In this interval, the best approximation of $f(y)$ (the minimum of $\int_0^4(f(y)-g(y)^2\mathrm{d}y$) is around $\alpha \simeq 1.7$. 
Therefore  $\tilde{k}$ is definitively different from the average connectivity $\langle k \rangle =2m$. 


\begin{figure}[t]
 \begin{center}
  \includegraphics[width=\columnwidth]{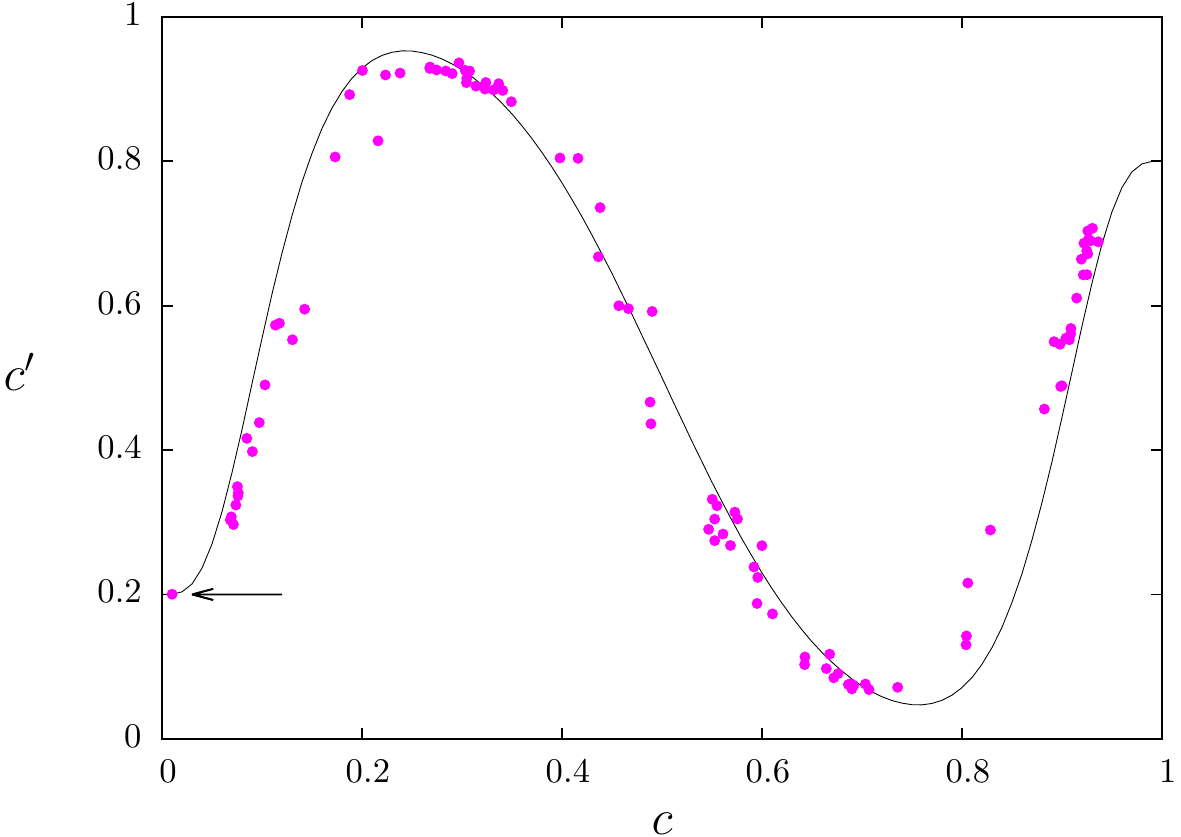}
 \end{center}
 \caption{\label{fig:rtn-sfn} (Color online)
  First 100 steps of the return map  for the density $c$ of the model
  on a scale free network with $N=10^4$, $m=20$,  $J=-4$. The first iterate is marked by the arrow. The continuous curve is the
  graph of Eq.~(\ref{mfK}) with $k=34$ }
\end{figure}

In conclusions, also in the case of a non-assortative scale-free network, the probability of getting a site with value 1 in the mean field approximation is given by 
\begin{equation}\label{mfK}
c' = \sum_{j=0}^{\tilde k} c^{j}\;(1-c)^{\tilde{k}-j}\;\tau\left(\frac{j}{\tilde{k}}\right),
\end{equation}
with $\tilde{k}\simeq 1.7 m$.

As usual, the mean field predictions are only approximately followed by actual simulations. 
In Fig.~\ref{fig:rtn-sfn}  we show the first 100 steps of the return map of the density $c$ for $J=-0.4$. The scale-free network is fixed, with $m=20$, $N=10,000$ and the initial opinions of the agents are chosen at random with $c=0.01$. The arrow marks the first point, that follows the mean field prediction with $\alpha=1.7$ ($\tilde k=34$), as in Fig.~\ref{fig:rtn-sfn}, 
but then, due to correlations, the return maps follows a different curve. This implies that nontrivial correlations establish also in  scale-free networks.

\end{document}